# IBEX: An open and extensible method for high content multiplex imaging of diverse tissues

Andrea J. Radtke[1,13,*], Colin J. Chu[2,3,13], Ziv Yaniv[4], Li Yao[5], James Marr[6], Rebecca T. Beuschel[1], Hiroshi Ichise[2], Anita Gola[2,7], Juraj Kabat[8], Bradley Lowekamp[4], Emily Speranza[2,9], Joshua Croteau[10], Nishant Thakur[2], Danny Jonigk[11], Jeremy Davis[12], Jonathan M. Hernandez[12], and Ronald N. Germain[1,2,*]

[1]Center for Advanced Tissue Imaging, Laboratory of Immune System Biology, NIAID, NIH, Bethesda, MD, USA

[2]Lymphocyte Biology Section, Laboratory of Immune System Biology, NIAID, NIH, Bethesda, MD, USA

[3]Translational Health Sciences, Bristol Medical School, University of Bristol, United Kingdom

[4]Bioinformatics and Computational Bioscience Branch, NIAID, NIH, Bethesda, MD, USA

[5]Howard Hughes Medical Institute, Chevy Chase, MD, USA

[6]Leica Microsystems Inc., Wetzlar, Germany

[7]Howard Hughes Medical Institute, Robin Neustein Laboratory of Mammalian Cell Biology and Development, The Rockefeller University, New York, NY, USA

[8]Biological Imaging Section, Research Technologies Branch, NIAID, NIH, Bethesda, MD, USA

[9]Innate Immunity and Pathogenesis Section, Laboratory of Virology, NIAID, NIH, Hamilton, MT, USA

[10]Department of Business Development, BioLegend, Inc., San Diego, CA, USA

[11]Institute of Pathology, Hannover Medical School, Hannover, Germany, Member of the German Center for Lung Research (DZL), Biomedical Research in Endstage and Obstructive Lung Disease Hannover (BREATH)

[12]Surgical Oncology Program, Metastasis Biology Section, Center for Cancer Research, National Cancer Institute, NIH, Bethesda, MD, USA

[13]Co-first authors

[*]Co-corresponding authors: Andrea J. Radtke (andrea.radtke@nih.gov); Ronald N. Germain, (rgermain@niaid.nih.gov)

Andrea J. Radtke: 0000-0003-4379-8967, Colin J. Chu: 0000-0003-2088-8310, Ziv Yaniv: 0000-0003-0315-7727, Li Yao: 0000-0003-2390-083X, James Marr: 0000-0001-8353-1429, Rebecca T. Beuschel: 0000-0002-3882-457X, Hiroshi Ichise: 0000-0002-5187-810X, Anita Gola: 0000-0003-1431-1398, Juraj Kabat: 0000-0001-8636-542X, Bradley Lowekamp: 0000-0002-4579-5738, Emily Speranza: 0000-0003-0666-4804, Joshua Croteau: 0000-0002-8142-0482, Nishant Thakur: 0000-0003-3699-0037, Jeremy Davis: 0000-0002-6334-6936, and Ronald N. Germain: 0000-0003-1495-9143.

## Abstract

High content imaging is needed to catalogue the variety of cellular phenotypes and multi-cellular ecosystems present in metazoan tissues. We recently developed Iterative Bleaching Extends multi-pleXity (IBEX), an iterative immunolabeling and chemical bleaching method that enables multiplexed imaging (>65 parameters) in diverse tissues, including human organs relevant for international consortia efforts. IBEX is compatible with over 250 commercially available antibodies, 16 unique fluorophores, and can be easily adopted to different imaging platforms using slides and non-proprietary imaging chambers. The overall protocol consists of iterative cycles of antibody labelling, imaging, and chemical bleaching that can be completed at relatively low cost in 2-5 days by biologists with basic laboratory skills. To support widespread adoption, we provide extensive details on tissue processing, curated lists of validated antibodies, and tissue-specific panels for multiplex imaging. Furthermore, instructions are included on how to automate the method using competitively priced instruments and reagents. Finally, we present a software solution for image alignment that can be executed by individuals without programming experience using open source software and freeware. In summary, IBEX is an open and versatile method that can be readily implemented by academic laboratories and scaled to achieve high content mapping of diverse tissues in support of a Human Reference Atlas or other such applications.

## Introduction

Ambitious efforts across multiple consortia, including the Human Cell Atlas (HCA)[1] and Human Biomolecular Atlas Program (HuBMAP)[2], aim to characterize all cell types in the human body, with the ultimate goal of creating a Human Reference Atlas. While it is impossible to precisely estimate the number of cell types present in the human body, nor the complete list of biomarkers needed to define these unique entities, a recent report identified 1,534 anatomical structures, 622 cell types, and 2,154 biomarkers (632 of which were proteins) in 11 human organs[3]. To provide a spatial context for these complex data, the field needs high content methods that capture the *in situ* biology of diverse tissues with sufficient coverage depth. Beyond supporting efforts to build a comprehensive map of healthy human tissues, high content imaging is critical for understanding tumor-immune interactions as well as the histopathology of disease.

To this end, several multiplexed antibody-based imaging methods have been developed[4-15] and reviewed[16-18]. Furthermore, a detailed comparison of these methods can be found in a recent commentary authored by domain experts in the field of spatial proteomics[19]. The majority of these existing methods generate high dimensional datasets through an iterative multistep process (a cycle) that includes: i) immunolabeling with antibodies, ii) image acquisition, and iii) fluorophore inactivation or antibody/chromogen removal. In addition to fluorophore-labelled antibodies, alternative methods such as immunostaining with signal amplification by exchange reaction (Immuno-SABER)[13] and co-detection by indexing (CODEX)[12] employ antibody DNA-barcoding with complementary fluorescent oligonucleotides to collect multiplexed imaging data. Here, fluorescent oligonucleotides are rapidly hybridized and dehybridized to visualize multiple biomarkers *in situ*. In contrast to these cyclic imaging methods, technologies utilizing metal-conjugated antibodies, such as multiplexed ion beam imaging (MIBI)[14] and imaging mass cytometry (IMC)[15], probe >40 markers without iterative antibody labeling or removal steps. Although commercialization of CODEX, MIBI, and IMC have enabled deep spatial profiling of tissue samples from large clinical cohorts[20-22], these methods are limited to specialized instruments, proprietary reagents, and, for mass spectrometry (MS)-based methods, may require highly trained engineering staff for instrument support.

For these reasons, we developed Iterative Bleaching Extends multi-pleXity (IBEX)[23], a high content imaging method that can be implemented by scientists with basic laboratory skills at relatively low cost. Using commercially available reagents and microscopes, IBEX has been used to spatially characterize complex phenotypes in tissues from experimental animal models[24] as well as clinically relevant human samples[23]. Because antibody validation and panel development can be costly in terms of time and capital[19], we provide curated lists of antibodies as well as example panels for several tissues. The majority of these resources are optimized for a fixed frozen method of tissue preservation that overcomes the need for antigen retrieval; however, we demonstrate how IBEX can be adopted to formalin-fixed, paraffin embedded (FFPE) tissues. While not documented here, we have previously shown that the basic IBEX workflow enables multiplexed imaging of heavily fixed tissues using Opal fluorophores and is compatible with commercially available oligonucleotide-conjugated antibodies[23]. In this paper, we provide extensive details and representative data for execution of the IBEX protocol along with instructions on how to automate and accelerate data collection using a cost-effective fluidics device and widefield microscope. We additionally provide guidelines for the proper orientation and embedding of tissues to standardize data interpretation across specimens. To facilitate widespread adoption, we present a software solution for image alignment based on a simplified open source interface to the Insight Toolkit (SimpleITK)[25,26] that can be used by laboratory scientists without programming experience.

**Development of the protocol**

IBEX expands on earlier work demonstrating the feasibility of using borohydride derivates to bleach fluorescently conjugated antibodies for multiplex imaging[9,27,28]. Bolognesi *et al.* and others have provided details on the chemistry, fluorophore inactivation, and proteolytic properties of sodium borohydride under their user-defined experimental conditions[9,27-33]. For the manual IBEX method described here, we consistently found that 15 minutes of exposure to 1 mg/ml of Lithium Borohydride ($LiBH_4$), a strong reducing agent[34], eliminated fluorescence signal from the following dyes: Pacific Blue, Alexa Fluor (AF)488, AF532, Phycoerythrin (PE), AF555, eFluor (eF)570, iFluor (iF)594, AF647, eF660, AF680, AF700, and AF750. Other fluorophores (Fluorescein isothiocyanate (FITC)) required a longer incubation time (<30 minutes) or bleaching in the presence of light (Brilliant Violet (BV) BV421 and BV510) for signal loss. Finally, certain fluorophores (Hoechst, JOJO-1, AF594, eF615) maintained their signal over multiple bleaching and imaging cycles (Table S1). The new automated IBEX protocol described in this report requires 0.5 mg/ml of $LiBH_4$ treatment for 10 minutes or, in some cases, 20 minutes under constant flow (120 µl/minute) for fluorophore inactivation. The continual exchange of fresh 0.5 mg/ml of $LiBH_4$ (automated), as opposed to a one-time application of 1 mg/ml (manual), enables efficient bleaching at a reduced concentration of $LiBH_4$ while preventing excessive bubble formation within the closed bath chamber, an issue not encountered with the manual method.

Tissue-specific panels are designed using antibodies conjugated to $LiBH_4$-sensitive dyes with a $LiBH_4$-resistant dye (usually Hoechst) serving as a fiducial for image alignment across iterative cycles. In our original publication we demonstrated that $LiBH_4$ acts by eliminating fluorescence signal by fluorophore inactivation and not antibody removal[23]. This feature allows for the inclusion of unconjugated primary antibodies in earlier cycles and amplification with fluorescently conjugated secondary antibodies in later cycles as outlined in the automated IBEX panels deposited here (Table S2). While both IBEX methods are compatible with indirect immunolabeling, we prefer to use directly labeled antibodies from vendors or conjugated in house using commercial kits. In addition to the points raised by Hickey and colleagues[19], we recommend purchasing >100 µg of antibody in a suitable format for conjugation; concentration of 1 mg/ml or greater in a buffer without carrier proteins or additives. For panels requiring multiple unconjugated antibodies, we advise placing these antibodies in earlier cycles to avoid unwanted detection by the secondary antibody with directly conjugated antibodies generated in the same host. Furthermore, primary antibodies must be raised in different hosts (rabbit, goat, and rat) versus multiple clones from the same host (3 rabbit clones) when using unconjugated versions with secondary antibody detection. Lastly, the selected secondary antibodies should be highly cross-adsorbed against antibodies from other host species. Here, we provide additional details for pairing antibodies and fluorophores for optimal imaging using manual and automated IBEX methods (Tables S3). Importantly, the number of parameters that can be imaged per IBEX cycle is dependent on several variables including microscope configuration, tissue autofluorescence, availability of conjugated antibodies, and overall panel design (Tables S2-S3).

The resulting method, IBEX, provides an efficient means for evaluating the spatial landscape of diverse tissues by offering: i) an adhesive that securely attaches tissues to the slide/coverslip surface over multiple fluid exchanges, ii) an abbreviated antibody labelling step to shorten the overall protocol, and iii) an open source image registration workflow. Using the tissue adhesive chrome gelatin alum, IBEX can be performed on the most delicate of tissues without tissue loss or degradation for up to 20 cycles in some cases[23]. Antibody labelling is accelerated using a specialized non-heating microwave for the manual IBEX method (30-45 minutes per cycle) and microscope stage heater at 37°C for the automated IBEX method (1 hour per cycle). The SimpleITK workflow provides registration of images acquired as a 3D tissue stack (manual IBEX) or single 2D z-slice (automated IBEX). We document an easy-to-use Imaris XTension (Bitplane) for registration that can be executed by individuals without programming experience. As an additional resource, an overview of costs is provided to aid potential users in writing an instrument grant application or adopting IBEX in their laboratory (Table S4).

**Overview of the protocol**

Here, we provide detailed instructions for obtaining high content images from a wide range of tissues using human material as exemplars (Figure 1). The protocol outlines tissue grossing (Steps 1-2), tissue processing (Steps 3-7), manual (Steps 8-38) or automated (Steps 39-77) IBEX imaging, and image alignment with SimpleITK (Steps 78-90). The overall workflow consists of 1) specimen preparation and preservation, 2) IBEX imaging, and 3) image processing that can be executed in 2-5 days or paused after each stage.

**Applications**

The target audience for IBEX ranges from academic laboratories desiring an affordable method for highly multiplexed imaging to commercial entities seeking a scalable workflow for characterizing clinical samples. Published studies have evaluated the distribution of immune cells in the livers of prime and target vaccinated animals[24] and visualized diverse stromal populations in human and mouse tissues using multiple markers[23]. Ongoing studies include, but are not limited, to assessing the tumor microenvironment of clinical samples and profiling unique epithelial cell populations present within the human thymus. The additional advancements described here—automation and an easy-to-use image alignment software package—make it an attractive method to gain additional insight into the cellular ecosystems of diverse tissues.

**Experimental design**

*Tissue processing*

In general, sample collection informed by anatomical landmarks (e.g., right axillary lymph node) and histological features (e.g. spatially invariant vasculature) is critical for proper specimen evaluation[35]. For tissue mapping efforts in support of a Human Reference Atlas, precise anatomical locations are required to integrate molecular and spatial data across individuals using a common coordinate framework (CCF)[2,36]. For these reasons, it is important to record these details and, if possible, discuss with the contributing surgeon, radiologist, and pathologist if unclear. Because the process of autolysis begins immediately after surgical removal of tissues or biopsy, we recommend placing specimens in fixative as soon as possible to prevent tissue degradation and thus preserve immunoreactivity and cytomorphology[35,37]. The most standard fixative, 10% neutral buffered formalin (4% formaldehyde), fixes tissue at a rate of 1-2 mm per hour at room temperature[37,38]. To ensure appropriate fixation, tissues should be cut into thin sections (less than 3-5 mm thick), placed in 15-20 fold the volume of fresh fixative, and incubated for a sufficient length of time (12-20 hours)[37,38]. IBEX is compatible with FFPE samples and detailed protocols for the preparation of such samples are outlined elsewhere[38]. To overcome the technical challenges posed by FFPE samples including epitope loss, the need for antigen retrieval, and high autofluorescence, we utilize a fixative with ~1% (vol/vol) formaldehyde and a gentle detergent. Following fixation, samples are immersed in 30% (wt/vol) sucrose for cryoprotection. Using this fixed frozen method, our laboratory has performed multiplexed imaging on a wide range of murine and human tissues[39-46]. Beyond excellent tissue preservation, this protocol significantly reduces autofluorescence by lysing red blood cells and reducing lipid-containing pigments. Prior to tissue embedding, it is important to properly orient specimens using guidelines established by surgical pathologists[35,37]. We provide detailed descriptions in the Procedure section for optimal handling of the human bowel, skin, lymph node, and wedge resections (liver, kidney, spleen) as examples (Box 1). We are presently finalizing panels and developing workflows for a greater range of tissues including human and mouse lung, thymus, and various tumors, with our current data showing all these tissues to be compatible with the IBEX protocol.

*Antibody validation and panel design*

A central tenet of IBEX, and other such methods[4-9,11-15], is the collection of reproducible multiplexed imaging data using antibodies specific for their intended targets. Recently, members of multiple tissue-mapping consortia including the HCA[1], HuBMAP[2], Human Tumor Atlas Network (HTAN)[47], and Human Protein Atlas[48], established guidelines for multiplexed antibody-based imaging methods[19]. We refer the reader to this work as it documents how to properly validate antibodies, build and test multiplexed antibody panels, conjugate custom antibody reagents, and overcome tissue autofluorescence. Furthermore, this work provides lists of antibody clones (>400 human; >90 mouse) that have been benchmarked by different users across distinct imaging platforms, including IBEX. An additional resource for qualifying multiplexed imaging panels can be found in a detailed protocol authored by Du and colleagues using tissue-based cyclic immunofluorescence (t-CyCIF)[49]. Of note, this work provides several considerations for characterizing antibody immunolabeling at the pixel-, cell-, and tissue-level and offers guidelines for minimizing artifacts arising from iterative imaging methods. Beyond these technical considerations, skillful panel design requires clarity on the scientific questions to be addressed as well as knowledge of the anatomical structures and cell types present within a given organ. The anatomical structures, cell types, plus biomarkers (ASCT+B) tables—assembled by more than 50 domain experts—are a valuable resource for understanding the tissue microanatomy of 11 human organs (Bone marrow and blood, Brain, Heart, Large intestine, Kidney, Lung, Lymph node, Skin, Spleen, Thymus, Vasculature) with more in the pipeline[3]. Furthermore, one can explore these data with a state-of-the-art visualization tool, aiding in the selection of cell-specific protein biomarkers (BP) to target for multiplexed imaging (https://hubmapconsortium.github.io/ccf-asct-reporter/).

Besides these resources, we provide antibodies, multiplexed tissue panels, and several IBEX-specific guidelines to aid in high quality data generation (Tables S1-S3). Because immunolabeling can vary across tissues, fixation conditions, and the imaging system employed, we recommend careful titration of all antibodies prior to IBEX imaging. Additionally, it is important to test whether novel antibodies are sensitive to $LiBH_4$ before adding them to established multi-cycle panels. This can be determined by comparing immunolabeling in serial sections with or without $LiBH_4$ pre-treatment. For evaluating potential epitope loss resulting from steric hindrance, we compare the spatial distribution patterns of antibody panels acquired serially to ones acquired iteratively via IBEX as we have previously shown[23]. In rare instances, if a particular antibody is thought to be impacted by $LiBH_4$ exposure or cycle order, the affected antibody is moved to an earlier panel. To elaborate on the effect of cycle number on antigenicity, antibodies directed against CD106 (RRID: AB_314561), Chromogranin A (RRID: AB_2892553), and DCAMKL1 (RRID: AB_873537) performed better in cycle 1 than in cycles 10, 7, and 5, respectively, in a 48-plex human thymus imaging panel (unpublished work). This issue impacts <2% of IBEX characterized antibodies versus the 12%[6] to 15-20%[8] of antibodies shown to be affected by cycle number in methods employing hydrogen peroxide ($H_2O_2$) as a fluorophore inactivation agent. However, we have observed diminished signal intensity with AF700-conjugated antibodies placed in later cycles (after cycle 4) as opposed to the same clone conjugated to other fluorophores. This, coupled with the limited availability of AF700-conjugated antibodies, reduces the number of parameters that can be robustly imaged over successive cycles. Given the importance of fluorophore conjugate on antibody performance, we provide guidelines for antibody-fluorophore pairing specific to manual and automated modes of IBEX (Tables S3).

*Manual IBEX*

The manual IBEX method is compatible with a wide range of imaging substrates and can be easily adapted to diverse systems including upright and inverted microscopes. We have performed IBEX on upright microscopes with tissues adhered to slides. However, we prefer using an inverted microscope with a chambered coverglass as it eliminates the need for coverslip removal between cycles. When preparing chambered coverglass samples, it is important to place the tissue in the center of the well to prevent any damage to microscope objectives. To ensure a uniform focal plane and secure adherence, the tissue must be carefully flattened onto the glass surface with a paintbrush (Extended Data Figure 1a). For 20-30 µm tissue sections, we utilize the PELCO BioWave Pro 36500-230 microwave for antibody labelling; however, comparable immunolabeling is observed for thin (~5-10 µm) tissue sections incubated for 1 hour at 37°C. Before fluorophore inactivation with $LiBH_4$, we extensively wash the sample (3 exchanges of 1 ml of PBS) to remove mounting media. We additionally wait until small bubbles form in the $LiBH_4$ solution, typically 10 minutes after dissolving in water, before adding to the samples. Small bubbles routinely form on the tissue during the 15-minute incubation of $LiBH_4$ (Extended Data Figure 1b). Because the manual IBEX method involves removing the sample from the microscope stage for iterative rounds of fluorophore inactivation and immunolabeling, careful attention must be paid to the alignment of images acquired over distinct cycles. We achieve this by matching shared landmarks, such as distinct nuclear morphology, in the first z-slice (Begin) and last z-slice (End) across the different image volumes (Extended Data Figure 1c). Manual execution is required in instances where a fluidics device is not available or compatible with existing instrumentation. Indeed, the IBEX method was initially developed using an advanced confocal instrument equipped with a 405 nm laser and a white light laser source producing a continuous spectral output between 470 and 670 nm. This system confers several advantages over conventional widefield microscopes including the ability to image more markers per cycle and the acquisition of fully resolved 3D z-stacks. However, this instrument is presently incompatible with the automated workflow described below. In addition to overcoming instrument limitations, the manual method provides users a means to acquire highly multiplexed images under experimental conditions not amenable to IBEX automation, e.g. access to pre-cut FFPE slides only.

*Automated IBEX*

To enable widespread adoption and higher throughput, we designed an automated workflow that could be easily implemented by most laboratories and possessed an economical equipment footprint (Figure 2a-c, Extended Data Figure 2). We achieved this goal by integrating an inverted widefield microscope (Leica Microsystems THUNDER imaging system) and a compact fluidics device (Fluigent ARIA). Our selection of an imager was based on the following technical specifications: i) ability to image multiple fluorophores per cycle, ii) stable multi-hour (>16 hours) acquisitions with pixel-pixel alignment across imaging cycles, iii) external triggering capabilities to interface with fluidics device, and iv) capture of multiple regions of interest with tiling in 2D and 3D. Of note, the THUNDER computational clearing and adaptive deconvolution software provides confocal quality images (based on contrast) in a fraction of the time and cost of traditional laser scanning confocal microscopes, making it particularly attractive for IBEX automation. With regards to the fluidics device, we required an instrument that was able to send and receive Transistor-Transistor Logic (TTL) pulses while supporting delivery of multiple solutions. The Fluigent ARIA possesses 10 reservoirs and typical experiments utilize the following configuration: i) $LiBH_4$ in Reservoir 1, ii) Hoechst labelling in Reservoir 2, iii) Antibody panels in Reservoirs 3-8, and iv) PBS wash in Reservoir 10 with an empty Reservoir 9. However, it is possible to deliver 7 unique antibody solutions with 4 fluorophores per cycle to achieve a 29-parameter dataset (7 cycles of 4 fluorophores per cycle plus Hoechst as a fiducial) in one execution of the program. Samples are prepared using coated 22 mm square coverslips and the RC-21B closed bath imaging chamber supplied by Warner Instruments (Figure 2b, Extended Data Figure 2). In addition to supporting linear solution flow, the closed bath system provides a large viewing area with a total bath volume of 358 μl, greatly reducing antibody labelling costs. Together, the protocol steps of immunolabeling, imaging, and fluorophore inactivation (Figure 2c) can be automated using commercially available instruments and reagents to obtain highly multiplexed images.

*SimpleITK image registration*

Both versions of IBEX yield a series of images that are collected separately as either 3D image stacks (manual) or 2D single z-slices (automated). Using SimpleITK, we developed an intensity-based image alignment protocol to register 2D and 3D IBEX-generated images. In our initial report, we improved upon existing algorithms[50,51] by providing a workflow capable of aligning large datasets (>260 GB) that additionally offers flexibility with regard to the fiducial used for registration[23]. To assess the quality of image alignment, a cross correlation matrix may be generated using a marker channel that is repeated across the cycles. Here, we document a user-friendly solution for image alignment by developing an executable that is compatible with Imaris, a commercial microscopy analysis software suite in worldwide use. Through the freeware Imaris Viewer and SimpleITK Imaris XTension, biologists without programming experience can obtain cell-cell registration across x-y-z dimensions from iterative cycles of IBEX at limited cost.

**Advantages and limitations**

IBEX has several advantages over existing multiplexed antibody-based imaging methods including its i) open and flexible nature, ii) capacity to utilize different imaging platforms, iii) short immunolabeling and bleaching time, iv) unrestricted compatibility with hundreds of commercial antibodies conjugated to diverse fluorophores, and v) simple implementation by biologists in standard laboratory environments. To date, we have demonstrated that 12 fluorophores, corresponding to 8 distinct imaging channels, can be inactivated within 10-20 minutes of $LiBH_4$ treatment alone. In contrast, methods employing $H_2O_2$ for fluorophore inactivation are limited to 7 fluorophores, representing 3 imaging channels, and require 1 hour in the presence of light for signal elimination[8]. Thus, the ability to inactivate diverse fluorophores in a short period of time is a significant advantage over existing iterative imaging methods. Importantly, treatment of fixed frozen tissue sections with $LiBH_4$ does not result in tissue loss as we previously found for $H_2O_2$[23]. The work outlined here and previously[23] offer a resource to the field of antibody-based imaging by providing exhaustive lists of validated reagents and approved panels for mouse and human tissues. It is our intent to enable high quality data generation while defraying the substantial cost, many thousands of US dollars for a 50-plex panel, and time, around 6 months, to develop equivalent panels *de novo*[19]. The primary limitation of the original IBEX method was that it required manual execution; however, we now have an affordable automated solution for high content imaging (Table S4). Presently, the thickness of tissue that can be imaged is restricted to <50 μm due to local tissue warping and distortion across iterative cycles. Beyond increasing the tissue volume that can be acquired (>400 μm), future improvements include custom fabrication of imaging chambers >22 $mm^2$ to enable automated acquisition of larger tissue areas including whole slide scans and tissue microarrays. We are actively working on methods that increase the number of parameters that can be imaged in a single IBEX cycle by employing additional IBEX compatible fluorophores and confocal microscopes equipped with advanced detectors and extended spectral outputs of 440-790 nm. While high numerical aperture (NA) objectives (40x, 1.3) have been shown to work well in IBEX imaging, we have not yet carefully tested higher magnification (63x, 100x) or higher NA (1.4) objectives for resolving subcellular structures.

## Materials

- *Ethics and biological materials*
  - Human Tissue
    - Human tissue was obtained on a NIH Institutional Review Board (IRB)-approved protocol (13-C-0076) at the time of risk-reducing surgery performed as a consequence of germline genetic mutation(s). All tissue procured, which included biopsies of lymph nodes, skin, spleen, liver and jejunum, was grossly normal as determined by the operative surgeon and histopathologically normal as determined by a board-certified pathologist. Of note, all tissue was obtained within twenty minutes of skin incision given our observation of neutrophil infiltrate with prolonged (>1 hour) procedures. Human kidney samples were collected from patients undergoing elective renal surgery at Hannover Medical School. Samples were enrolled in this study after histologic assessment only after completion of routine diagnostics and written consent approved by the local ethics committee of Hannover Medical School (ethics-vote number: 3381-16, 2893-15, 1741-13).
    - Caution: Any experiments involving human tissues must conform to relevant institutional and national regulations. All procedures described in this protocol were approved by an IRB.

## Reagents and consumables

- *Tissue grossing and processing*
  - BD Cytofix/Cytoperm (BD Biosciences, cat no. 554722)
  - Camel hair brush (Ted Pella, Inc., cat no. 11859)
  - Cell culture plate 6 well (Corning, cat no. 3335)
  - Cell Pro 500 ml 0.22 µm Bottle Top Filter (Alkali Scientific, cat no. VH50022)
  - Histomolds, 10mm x 10mm x 5mm (Sakura, cat no. 4565)
  - Histomolds, 15mm x 15mm x 5mm (Sakura, cat no. 4566)
  - Low profile microtome blades (Leica Biosystems, cat no. 14035843496)
  - Optimal cutting temperature (OCT) compound (Sakura, cat no. 4583)
  - Regular bevel needles 30G (BD, cat no. 305106)
  - Sucrose Solution
    - Prepare by dissolving 30% (wt/vol) Sucrose (Millipore Sigma, cat no. S0389) in sterile 1X PBS and pass through a 0.22 µm bottle top filter. The solution can be stored at 4°C for several months if kept sterile.
  - UltraPure agarose (ThermoFisher Scientific, cat no. 16500-500)

- *Manual and automated IBEX: Tissue blocking, antibody immunolabeling, and fluorophore inactivation*
    - Antibodies (See Tables S1-S2)
    - Blocking Buffer
      - Solution made from 1X PBS (Gibco, cat no. 10010-023) containing 0.3% Triton-X-100 (Millipore Sigma, cat no. 93443), 1% Bovine Serum Albumin (Millipore Sigma, cat no. A1933) and 1:100 dilution of Human (BD, cat no. 564219) or Mouse (BD, cat no. 553141) Fc-block
    - di$H_2O$ (Quality Biological, cat no. 351-029-101)
    - Fluoromount-G (Southern Biotech, cat no. 0100-01)
    - Hoechst 33342 (ThermoFisher Scientific, cat no. H3570)
    - Lithium borohydride (STREM Chemicals, cat no. 93-0397)
    - 20 ml Disposable syringe with luer-lock tip (EXEL Int., cat no. 26280)

- Millex-GS Syringe Filter Unit (0.22 µm) (Millipore Sigma, cat no. SLGSM33SS)

- *Manual IBEX*
  - *Sample preparation (Fixed frozen)*
    - Chrome Alum Gelatin (Newcomer Supply, cat no. 1033A)
    - 2-well chambered coverglass (Lab-Tek, cat no. 155380)

- *Automated IBEX*
  - *Sample preparation (Fixed frozen)*
    - Chrome Alum Gelatin (Newcomer Supply, cat no. 1033A)
    - Cover glass 22x22 mm #1.5 (Electron Microscopy Sciences, cat no. 63786-10
  - *Sample preparation (FFPE)*
    - AR6 buffer 10X (Akoya Biosciences, cat no. AR600250ML)
    - Chrome Alum Gelatin (Newcomer Supply, cat no. 1033A)
    - Cover glass 22x22 mm #1.5 (Electron Microscopy Sciences, cat no. 63786-10
    - Ethanol, 200 Proof (Decon Labs, Inc., cat no. 2701)
    - Formalin, 10% neutral buffered (Cancer Diagnostics, Inc., cat no. FX1003)
    - TBST wash buffer
      - Solution made from 1X TBS (Quality Biological, cat no. 351-086-101) and 0.05% Tween 20 in di$H_2O$ (Millipore Sigma, cat no. 9005-64-5)
    - Xylene, histology grade (Newcomer Supply, cat no. 1446C)
  - *Chamber assembly*
    - Dow Corning® high-vacuum silicone grease (Millipore Sigma, cat no. Z273554)
    - 10 ml size Luer-lock syringe (BD, cat no. 309604)

**Equipment and tools**

- *Tissue grossing, processing, and sample preparation*
  - Cryostat (Leica Biosystems, CM1950) or a comparable instrument
  - Dissecting mat, flexible, polypropylene (Newcomer Supply, cat no. 5218A)
  - Dissecting needles (Newcomer Supply, cat no. 5220PL)
  - Fine forceps (Fine Science Tools, cat no. 11412-11)
  - Forceps, custom embedding 13 cm, curved, standard grade (Newcomer Supply, cat no. 5536)
  - Scissors (Fine Science Tools, cat no. 114090-09)
  - Small digital incubator (Boekel Scientific, 133000)
  - Sterile disposable scalpels #11 (Newcomer Supply, cat no. 6802A)
  - Stereomicroscope (Zeiss, Stemi 2000-CS) or a comparable instrument
  - Stereomicroscope illuminator (Zeiss, KL 1500 LCD) or a comparable instrument

- *Automated IBEX: Sample preparation (FFPE)*
  - EasyDip slide staining kit (Newcomer Supply, cat no. 5300KIT)
  - EasyDip anodized aluminum jar rack holder (Newcomer Supply, cat no. 5300JRK)
  - Glass beaker, 100 ml (VWR, cat no. 10754-948)
  - Rotary microtome (Leica Biosystems, RM2255) or a comparable instrument
  - StatMark pen (Electron Microscopy Sciences, cat no. 72109-12)
  - Wash N'Dry cover slip rack (Electron Microscopy Sciences, cat no. 70366-16)

- *Manual IBEX: Immunolabeling and microscopy*
    - Confocal microscope
        - The default for this protocol uses an inverted Leica TCS SP8 X confocal microscope equipped with a 40x/1.3 objective, 4 HyD and 1 PMT detectors, a white light laser source that produces a continuous spectral output between 470 and 670 nm as well as an additional 405 nm laser. Images were captured at 8-bit depth, with a line average of 3, and 1024x1024 format with the following pixel dimensions: x (0.284 µm), y (0.284 µm), and z (1 µm). Images were tiled and merged using the LAS X Navigator software (LAS X 3.5.5.19976). Other confocal microscopes with comparable specifications in terms of laser lines, objectives, detectors, precision stage, and control software can be used. We prefer inverted over upright microscopes because these systems eliminate the need for coverslip removal, a tedious step that can lead to tissue loss.
    - PELCO BioWave Pro microwave (Ted Pella Inc., 36500-230) equipped with a PELCO SteadyTemp Pro (Ted Pella Inc., 50062) thermoelectric recirculating chiller
        - This specialized, non-heating microwave accelerates immunolabeling without tissue loss or epitope damage.
    - Slide moisture chamber (Scientific Device Laboratory, cat no. 197-BL)

- *Automated IBEX: Immunolabeling and microscopy*
    - 31G insulin syringes (BD, cat no. 328438)
    - ARIA fluidics system
        - ARIA automated perfusion system (Fluigent, cat no. CB-SY-AR-01)
        - Fluigent Low Pressure Generator (FLPG) pressure supply (Fluigent, cat no. FLPG005)
    - BNC Male to SMB Plug Cable RG-316 Coax in 120-inch length (2 needed in total) for TTL triggering (Fairview Microwave, cat no. FMC0816315-120)
    - Imaging chamber, vacuum, and heater
        - CC-28 Cable assembly for heater controllers to platform (Warner Instruments, cat no. 64-0106)
        - Dedicated Workstation Vacuum System (Warner Instruments, cat no. 64-1940)
        - PM-2 Platform for Series 20 chambers, magnetic clamps, heated (Warner Instruments, cat no. 64-1561)
            - Before first use, the electrode prongs need to be carefully bent upwards 30 degrees to allow the assembled unit to fit into the stage adapter. This can be done with a pair of forceps.
        - RC-21B Large Closed Bath Imaging Chamber (Warner Instruments, cat no. 64-0224)
        - SA-20PL Series 20 stage adapter (Warner Instruments, cat no. 64-0299)
        - TC-324C Single Channel Temperature Controller (Warner Instruments, cat no. 64-2400)
    - Leica THUNDER microscope
        - The system employed in this automated IBEX protocol is a THUNDER 3D Cell Culture microscope with a high precision (Quantum) stage, a high quantum efficiency sCMOS camera (DFC9000 GTC) and a 40x/1.3 NA oil objective. The system includes an adaptive focus control unit, to ensure image focus over many hours of imaging and fluidic cycle times. The LED8 light source has 8 individual LED lines for excitation with millisecond triggering. Two advanced sequencer boards were installed in the control box of the microscope configured to use SMB connections to send and receive external triggers from additional

devices. A spill guard was installed on the objective turret to ensure no liquids enter the microscope body.

- For fluorescence imaging, a custom quad-band filter with external filter wheel (PN: 11536075) with two additional single-band filters (PN: 8118215) were used to image 7 dye channels per pass. The filter excitation, dichroic and emission lines are listed below.
  - Quad-Band cube: dichroic at 391/32, 479/33, 554/24, 638/31, no excitation or emission filters.
  - External filter wheel position 1: 434/32, position 2: 520/40, position 3: 585/20, position 4: 720/60, position 5: pass-through.
  - Single-band 1 - 585/22 excitation, 594 dichroic, 625/30 emission.
  - Single-band 2 - 635/20 excitation, 647 dichroic, 667/30 emission.
- LAS X [3.7.1.21655], the software used on the THUNDER imager, comes with the primary modules needed to perform the imaging shown here, including the THUNDER computational clearing and adaptive deconvolution routines that utilize Leica Microsystems proprietary algorithms to enhance image contrast and resolution. The additional 'Trigger to Peripherals' (PN: 11640613) module is required and the 'Dye Finder' (PN: 11640863) module is recommended to negate spectral cross-talk. All images were captured at a 16-bit depth with the following pixel dimensions: x (0.160 μm), y (0.160 μm), and z (1 μm).
- Other inverted widefield microscopes with comparable specifications in terms of light source and filter cubes for excitation and emission of multiple fluorophores, precision stage, adaptive focus control or other means of maintaining the same focus position over multiple cycles, objective choice, camera sensitivity and resolution, control software, deconvolution algorithms, and the ability to send and receive external triggers can be used.

- *Software*
  - Imaris and Imaris File Converter (x64, version 9.5.0 and higher, Bitplane)
  - Python (version 3.7.0 and higher)
  - Custom Imaris Extension (XTRegisterSameChannel, open source software distributed under Apache 2.0 license)

- *Equipment Setup*
  - <u>Automated IBEX connection of ARIA fluidics device to LAS X software</u>
    - Prior to using the ARIA with LAS X, the triggers must be programmed into your hardware configuration and the cabling attached. Here we name them 'ARIA to Leica' and 'Leica to ARIA'.
    - Before imaging, set up the triggers in the hardware configuration. Within LAS X hardware configurator, navigate to the sequencer tab under 'Configure'. Add two new triggers from the available triggers, if you do not have an open trigger on the sequencer you will need to a) add a sequencer or b) remove a triggered device.
    - For the first trigger, give it a name (i.e. 'Leica to ARIA'), select 'output' and list it as a TTL output. The second trigger will also be named (i.e. 'ARIA to Leica') and listed as a TTL but will be set as 'input'.
    - Once programmed, connect a SMB to BNC cable from the sequencer output for the 'output' trigger to the 'IN' connection on the ARIA.
    - Connect a second SMB to BNC cable from the sequencer 'input' trigger to the 'OUT' connection of the ARIA.
  - <u>Software Installation for SimpleITK Imaris Python Extension</u>

- For initial setup, install Python 3.7.0 and download our Imaris extensions code repository as a zip file: (https://github.com/niaid/imaris_extensions/archive/refs/heads/main.zip). Installation instructions are available online: [https://github.com/niaid/imaris_extensions] and in the README.md file which is part of the zip file.
- Additional details can be found in the XTRegisterSameChannel SimpleITK Imaris Python Extension YouTube tutorial (https://youtu.be/rrCajI8jroE). To illustrate the usage of this software we provide sample data on Zenodo [https://doi.org/10.5281/zenodo.4632320].

## Procedure

### Manual and Automated IBEX

*Tissue grossing*
Timing: 2-3 hours, depending on the number and type of samples

1) Survey the organ specimens for distinct anatomical landmarks that may aid in tissue orientation, and, whenever possible, preserve these metadata. If biopsies are assessed, clinical imaging data (Computerized Tomography (CT) scan, ultrasound) of the biopsy procedure is helpful.
2) Prepare samples with dissecting tools (scalpel, forceps, scissors) on a dissecting mat or other clean, fiber-free surface. For small samples, a stereomicroscope and illuminator may be required for careful dissection (Figure 1). See Box 1 for tissue specific details.
    Critical: Work quickly to minimize tissue damage from autolysis.
    Caution: Be careful around cutting implements and properly dispose of sharps. Use proper personal protective equipment (PPE) to reduce exposure to potential blood- or air-borne pathogens.

*Tissue processing*
Timing: 2 days

3) Our preferred method of fixation uses 1 ml of BD Cytofix/Cytoperm for every 3 ml of 1X pH 7.4 PBS. Please consult published methods for preparing FFPE samples[38]. Transfer thin (a few mm) tissue sections (skin, lymph node, wedge resections) to a 50 ml conical tube with 20 ml of fixative. For the bowel, completely submerge the pinned tissue in fixative. The ideal ratio of fixative to tissue is 20:1. Incubate at 4°C for 16-24 hours.
    Critical: Be mindful of the expiration date published on the BD Cytofix/Cytoperm bottle. If working with potentially infectious samples, identify the fixation conditions that fully inactivate pathogen(s) before preserving samples as described above. For example, an internal biosafety review committee found that increasing the incubation period to 72 hours inactivated SARS-CoV-2 virus particles in mouse lung tissues. Importantly, tissue epitopes were well preserved under these conditions.
    Caution: Formaldehyde is a known carcinogen and environmental hazard. Wear PPE and dispose of waste properly.
4) Wash samples twice, less than 5 minutes per wash, in 1X PBS at room temperature to remove excess fixative.
5) Place sample in 30% sucrose (wt/vol) and incubate at 4°C for 16-24 hours.
6) Fill a Histomold with OCT medium and embed the tissue. The tissue should occupy less than 60% of the mold volume. Freeze by placing the Histomold on dry ice. Using forceps, ensure the tissue is correctly oriented in the Histomold for subsequent sectioning. For orientation guidance see Box 1.
7) Wrap in foil and store at -80°C.
    Pause Point: Cryopreserved tissues can be left at -80°C for several years.

*Manual IBEX*

*Sample preparation*

Timing: 2.5 hours

8) Remove the tissue block from -80°C and allow it to equilibrate to the chamber temperature of the cryostat for 1 hour.
9) While blocks are equilibrating in the cryostat (1 hour), coat a 2-well chambered coverglass by pipetting 15 µl of chrome alum gelatin into each glass well. Use a pipette tip or smaller coverslip (18x18 mm) to spread the solution. Dry for 1 hour in a 60°C oven.

    Critical: It may require multiple passes to get an even film on the coverglass. Aspirate any large accumulations as these can form autofluorescent regions. Do not use the adhesive past the stated expiration date as tissue adherence will be compromised.
10) Prepare 20-30 µm sections with a cryostat. Section tissue directly onto the 2-well chambered coverglass.

    Critical: The tissue section should be completely flat and centered within the well. It is worth preparing several backup samples during each session.

    Caution: Be careful around cutting implements and properly dispose of sharps. Use PPE to reduce exposure to blood- or air-borne pathogens.
11) Dry sections onto coated 2-well chambered coverglass for 1 hour at 37°C or overnight at room temperature.

    Critical: Do not store chambered coverglass with sectioned tissue in a -20°C freezer or the tissue will lift. It is best to use the chambers within 1-2 days of preparation.

*Tissue blocking and antibody immunolabeling*

Timing: 1 hour

12) For a 2-well chambered coverglass, add 1 ml of 1X PBS to each well to rehydrate the tissue. Incubate for 5 minutes at room temperature.
13) Add 400 µl of Blocking Buffer per well. Place humidity chamber with 2-well chambered coverglass into the BioWave Pro microwave directly on the cooling plate. Switch on the SteadyTemp Pro 50062 Thermoelectric Recirculating Chiller set to 26°C.

    Critical: Ensure the cooling plate is filled with ultrapure water, no air or bubbles are present, and the temperature probe is secured. Make sure that the SteadyTemp cooler option is activated in the software.
14) Start one cycle of the microwave program. A 2-1-2-1-2-1-2-1-2 program is used where '2' denotes 2 minutes at 100 watts and '1' denotes 1 minute at 0 watts. The above program is executed once for blocking and secondary antibody labelling and twice for primary antibody labelling. Alternatively, thin tissue sections (5-10 µm) can be immunolabeled for 1 hour at 37°C.
15) Aspirate the Blocking Buffer.
16) Add 400 µl of primary antibody solution per well. Perform two cycles of the microwave program or incubate for 1 hour at 37°C (5-10 µm sections).

    Critical: It is important to carefully validate and titrate all antibodies for optimal immunolabeling prior to implementation in IBEX protocols, preferably with sections of the same tissue.
17) Aspirate the antibody solution and wash with 3 exchanges of 3 ml of 1X PBS. The entire wash step can be performed in 5 minutes.
18) If a secondary antibody is to be used add this now in 400 µl of Blocking Buffer and perform one cycle in the microwave as above or incubate for 1 hour at 37°C (5-10 µm sections). Then wash with 3 exchanges of 3 ml of 1X PBS.

    Critical: Do not include secondary antibodies that will cross-react with primary antibodies in multiplexed panels. Always use highly cross-adsorbed secondary antibodies to prevent off-target immunolabeling.
19) Add 1 ml of Fluoromount-G to samples.

    Critical: It is beneficial to leave samples for 30 minutes at room temperature in the dark to allow the tissue to equilibrate and settle before imaging.

Pause Point: Samples are stable in mounting media for several days at room temperature in the dark; however, we obtain the best results when images are acquired within 24 hours of immunolabeling.

*Microscope setup and imaging*
Timing: 4 hours

20) Clean the bottom of the chambered coverglass with lens cleaner and lens paper immediately before imaging. Add immersion oil to the glass surface and place on the microscope stage.
21) Slide the chamber firmly into the top left corner (or equivalent) of the stage insert, so the tissue will be consistently positioned in the XY plane.
22) Image on a confocal inverted microscope. We use a Leica TCS SP8 X system as outlined here.
23) Using Leica LAS X Navigator software, define the tissue volume to be acquired (number of tiles and z-stack size). Be sure to note the tile number and z-stack volume as this will be kept constant throughout all cycles.
24) To minimize spectral overlap, perform multiple scans using separate detectors with sequential laser excitation of compatible fluorophores.
25) Following image acquisition, correct for spectral overlap between fluorophores (Box 2).

*Fluorophore bleaching*
Timing: 30 minutes

26) Clean off immersion oil and thoroughly aspirate Fluoromount-G with a pipette.
27) Perform three exchanges of 3 mL of 1x PBS to completely dissolve and wash away the Fluoromount-G.
    Critical: It is imperative that all Fluoromount-G be removed for bleaching to be effective.
28) Prepare a 1 mg/mL solution of $LiBH_4$ in $diH_2O$.
    Caution: Perform these steps in a chemical hood with appropriate PPE. The reaction can generate hydrogen gas, which is flammable. To avoid flames, always work with small amounts (<10 mg) of $LiBH_4$. Store $LiBH_4$ in a sealed container with desiccant and use a new vial of $LiBH_4$ no later than 4 weeks after opening original container.
    Critical: Leave the solution to incubate at room temperature for 10 minutes after gentle mixing. Formation of bubbles indicates the solution is ready. The solution must be used within 4 hours or it becomes ineffective. The solution can contain manufacturing impurities and should be passed through a 0.22 µm syringe filter.
29) Add the $LiBH_4$ solution to the tissue. The tissue should be completely covered by the solution. For a chambered coverglass, we add 1 ml per well. Incubate for 10 minutes at room temperature exposed to standard ambient lighting.
30) Optional: If using BV421 and BV510 dyes, add $LiBH_4$ to the chamber and place on the microscope stage. Turn the epifluorescence lamp to maximum power, select the DAPI filter, focus on the tissue, and scan across the acquired region of interest using the eye piece to observe signal loss (1-2 minutes per field of view).
31) Perform three exchanges of 3 mL of 1X PBS to completely wash away the $LiBH_4$ solution.
    Critical: Incomplete washing can lead to bleaching of the subsequent panel of antibodies.

*Cycles of tissue labeling, image acquisition, and fluorophore bleaching*
Timing: 6 hours

32) Add the next panel of antibodies and repeat the procedure from Steps 16 to 19.
33) Extra attention is required when placing the sample back onto the microscope stage. The XY plane should be similar due to consistent positioning in the stage insert.
34) Open the image acquired from Cycle 1 in the LAS X Navigator software by right-clicking the desired image and selecting "Open in New Viewer". Scroll to the top of the z-stack (Begin) and click "Live" to view the sample presently on the microscope stage (Cycle 2, for example).

35) Using the channel that contains the repeated marker (Hoechst), identify unique structures (nuclear shapes) that are present in the "live" image (Cycle 2, for example) and the previous image (Cycle 1). Set this z-position as your Begin. To set the End, repeat this step while preserving the total z-stack volume captured in the previous image (8 µm).
36) Image using the same parameters established for the image acquired during Cycle 1.
    Critical: It may be necessary to check alignment across different XY regions of the tissue. Distinctive tissue structures such as nuclei are useful to achieve fine alignment (Extended Data Figure 1c).
37) Repeat Steps 26-31 to inactivate $LiBH_4$-sensitive fluorophores.
38) Repeat Steps 32-36 to reach the desired number of cycles and markers. Process images using Steps 78-90.

<u>*Automated IBEX*</u>

*Sample preparation*

Timing: 2 hours for fixed frozen tissues (40A); 3.5 hours for FFPE tissues (40B)

39) Coat a square 22x22 mm #1.5 glass coverslip by adding 3 µl of chrome alum gelatin to one side. Spread evenly using the edge of another coverslip and dry for 30-60 minutes in a 60°C oven. Denote the coated side of the coverslip using a symbol with a StatMark pen. We always mark the lower right corner with a series of characters that can only be read in one direction (QJZ). This prevents confusion if the coverslip is reoriented over the course of the experiment (sectioning, chamber assembly).
40) For fixed frozen tissue follow Option (A), for FFPE tissue follow Option (B):
    A. Fixed frozen tissue preparations
        I. Using a cryostat, section the OCT embedded tissue at 10 µm thickness onto the center of the coverslip and dry overnight at room temperature or for 1 hour at 37°C.
            Critical: Ensure the tissue section is completely flat and centered on the glass coverslip. It is worth preparing several backup samples during one session.
            Caution: Be careful around cutting implements and properly dispose of sharps. Use PPE to reduce exposure to blood- or air-borne pathogens.
            Pause Point: The dried samples can be stored (covered) at room temperature for up to 5 days before use; however, we recommend acquiring images within 1-2 days of sample preparation.
    B. FFPE tissue preparation
        I. Use a microtome to cut sections at 5 µm.
            Caution: Be careful around cutting implements and properly dispose of sharps. Use PPE to reduce exposure to blood- or air-borne pathogens.
        II. Float tissue onto coated coverslips using the same technique utilized for FFPE sections on glass slides.
        III. Place in a 60°C oven for 1 hour to adhere the tissues to the coverslips.
        IV. To deparaffinize the tissue, place coverslips in a Wash N'Dry coverslip rack. Add 100 ml of the solutions listed in (V-X) to individual jars (8 in total) of the EasyDip slide staining kit and rack. Use forceps to transfer the coverslip rack between jars. Do not allow the tissue to dry out at any stage.
        V. 100% Xylene for 10 minutes.
        VI. 100% Xylene for 10 minutes.
        VII. 100% Ethanol 10 minutes.
        VIII. 95% Ethanol for 10 minutes.
        IX. 70% Ethanol for 5 minutes with quick wash (< 1 minute) in water.
        X. 10% Formalin for 15 minutes with quick wash (< 1 minute) in water. TBST for 5 minutes.
        XI. For antigen retrieval, place coverslips (in Wash N'Dry coverslip rack) in a glass beaker with 100 ml of a 1X solution of AR6 buffer.

XII. Place beaker in a 95°C water bath and incubate for 40 minutes. Remove from the water bath and allow to gradually cool on the bench for at least 20 minutes.

XIII. Replace 1X AR6 buffer with 1X PBS and cover glass beaker with aluminum foil.

Pause Point: The samples can be stored in PBS at room temperature for up to 2 days before use.

Critical: Always check the antigen retrieval conditions specified by the antibody vendor before immunolabeling as the buffer(s) suitable for one epitope may be incompatible with another epitope. For this reason, we optimize the antigen retrieval protocol to obtain conditions that work for all antibodies used in our experiments.

Caution: Perform deparaffinization in a properly functioning chemical hood. Read the safety data sheets associated with these chemicals and take proper precautions before handling. Xylene and formalin (formaldehyde) are known physical and environmental hazards. Additionally, xylene and ethanol are highly flammable liquids.

*Chamber assembly*

Timing: 15 minutes

41) Fill a 10 ml Luer-lock syringe with vacuum grease, by initially removing and then replacing the plunger.

    Critical: Ensure the vacuum grease is fresh and white in color. Yellow discoloration is a sign of degradation and can lead to chamber leakage.

42) Coat the base of the RC-21B chamber first, by piping a line of vacuum grease 2 mm in diameter along the four sides of the square recession. Spread the grease evenly using a pipet tip.

    Critical: Avoid getting vacuum grease in the RC-21B fluid inlet or outlet ports to prevent fluid obstruction.

43) Apply the coverslip with adhered tissue to the base, ensuring the side with the tissue faces upwards into the chamber. Gently press around the perimeter of the coverslip to ensure a continuous interface of grease surrounds, but does not spill, into the central chamber.

    Critical: If using FFPE tissue, keep the tissue wet by adding a drop of PBS onto the tissue while preparing the chamber.

44) Repeat Steps 42 and 43 with the following alterations: coat the top (not the base) of the RC-21B chamber with vacuum grease and use a fresh 22x22 mm coverslip (no tissue) to create a closed bath chamber.

45) Place the RC-21B chamber into the PM-2 magnetic platform with outflow side adjacent to the Warner label. Apply the magnetic clamp, keeping the Warner label on the platform base proximal to the Warner label on the magnetic clamp. See Extended Data Figure 2.

*Prepare required solutions*

Timing: 45 minutes

46) PBS can be added directly from a 1X stock to the provided glass container in Reservoir 10 of the ARIA.

47) Hoechst and Blocking Solution: Add 1 µl of Hoechst 33342 and 10 µl Fc-block to 1 ml of Blocking Buffer described in Reagents section.

48) Antibody Panels: In 400 µl total volume of Blocking Buffer, prepare antibody panels for each iterative cycle using previously determined titrations.

    Critical: Do not include secondary antibodies that will cross-react with primary antibodies in multiplexed panels. Always use highly cross-adsorbed secondary antibodies to prevent off-target immunolabeling.

49) $LiBH_4$: Prepare at least 5 ml of a 0.5 mg/ml $LiBH_4$ solution (dissolved in di$H_2O$).

    Caution: Perform these steps in a chemical hood with appropriate PPE. The reaction can generate hydrogen gas which is flammable. To avoid flames, always work with small amounts (<10 mg) of $LiBH_4$. Store $LiBH_4$ with desiccant and use a new vial of $LiBH_4$ no more than 4 weeks after opening the primary container.

Critical: Only make this solution just before the run starts. The solution becomes less effective at bleaching and will need replacement after 4 hours. The solution can contain manufacturing impurities and must be passed through a 0.22 µm syringe filter before adding to the ARIA fluidics device.

*Prepare the ARIA unit*

Timing: 10 minutes

50) Switch on the FLPG unit and ensure a minimum of 2.2 bar is reached before switching on the ARIA unit power supply and starting the Fluigent Controller Software on the attached computer.

Critical: If not already done, perform a calibration run and record the values, making sure these recorded values appear whenever the software is restarted.

51) Create a custom ARIA program based on the example steps outlined below. Load this sequence before each run.

52) Add solutions to the reservoirs specified in the program on the ARIA machine. Our standard configuration is: $LiBH_4$ (Reservoir 1), Hoechst + Fc block (Reservoir 2), Antibody Panels (Reservoir 3-8), and PBS (Reservoir 10).

Critical: Only 15 ml Falcon™ brand tubes fit the taper to ensure an airtight seal.

Example ARIA Sequence Program for a 6 cycle experiment (Total time: ~630 minutes (10.5 hours)

| **Step** | **Step action** | **Flow rate (µl/min)** | **Total volume (µl)** | **Time (Minutes)** | **TTL** | **Purpose** |
|---|---|---|---|---|---|---|
| 0 | Prefill | | | 3 | | Load the fluid lines with PBS |
| 1 | Flush tubing | 200 | Variable | 0.5 | | Fill lines with PBS up to the 2-switch |
| 2 | Volume injection of PBS | 120 | 800 | ~7 | | Rehydrate tissue |
| 3 | Volume injection of Hoechst | 120 | 300 | 2.5 | | Nuclear labeling (Hoechst) |
| 4 | Wait | | | 5 | | Nuclear labeling with Hoechst |
| 5 | Volume injection of PBS | 120 | 600 | 5 | | PBS wash |
| 6 | Wait for user | | | 5-10 | | User defines ROI and z-slice |
| 7 | Volume injection of Antibody #1 | 120 | 300 | 2.5 | | Injection of Antibody |
| 8 | Wait | | | 60 | | Antibody immunolabeling |
| 9 | Volume injection of PBS | 120 | 1200 | 10 | ARIA to THUNDER | PBS wash and signal to THUNDER for imaging |
| 10 | Wait for TTL | | | ~10 | THUNDER to ARIA | Wait for imaging to complete and signal to ARIA |
| 11 | Volume injection of PBS | 120 | 240 | 2 | | Fluid spacer between antibodies and $LiBH_4$ |
| 12 | Volume injection of $LiBH_4$ | 120 | 1200 | 10 | | Fluorophore inactivation with $LiBH_4$ |
| 13 | Volume injection of PBS | 120 | 1200 | 10 | | PBS wash |
| 14 | Repeat steps 7-13 for 4 times | | | | | |
| 44 | Volume Injection | 120 | 300 | 2.5 | | Injection of Antibody #6 |
| 45 | Wait | | | 60 | | Antibody immunolabeling |
| 46 | Volume injection of PBS | 120 | 1200 | 10 | ARIA to THUNDER | PBS wash and signal to THUNDER for imaging |
| 47 | Wait for TTL | | | ~10 | THUNDER to ARIA | Wait for imaging to complete and signal to ARIA |
| 48 | Wait for user | | | Variable | | Instrument paused until user input received |

*Initial setup of the THUNDER microscope*
Timing: 5 minutes

53) Switch on microscope and LED8 source. Launch LAS X software. Select objective (this protocol is optimized for a 40x/1.30 oil immersion objective). Place the camera into 16-bit mode in the configuration tab. Select the required channels based on the fluorophores present in antibody panels. Set LED power to 25% with 150 ms exposure time for all.
    Critical: If using an AF750 filter combination, increase LED power to 40% for this channel as the signal is greatly reduced with the AFC function.
54) Highlight the Hoechst channel and turn on the Linked Shading feature. Follow the command prompts using the Leica THUNDER dialog box to set this for each objective on first use.
55) The LAS X Dye Separation module may be used to compensate for spectral spillover between channels. See Box 2 for details.

*Prepare microscope stage*
Timing: 5 minutes

56) Insert the Warner SA-20PL stage holder into the motorized microscope stage base.
57) Clean the bottom coverslip of the PM-2 thoroughly with lens cleaner to remove any vacuum grease or marks. If using the recommended 40x/1.30 oil objective, coat the bottom surface of the coverslip generously with immersion oil across a wide region and avoid bubbles.
58) Insert and fasten the assembled imaging chamber and PM-2 magnetic platform into the stage holder. Position the assembled unit so that the Warner name is on the right side and the baseplate temperature probe hole is at the top. See Extended Data Figure 2.
59) Attach the inlet tubing firmly into the RC-21B chamber.
60) Arrange and attach the electrical cables for the heater to the prongs on the PM-2 platform.
61) Insert the temperature probe into the hole at the top of the PM-2 platform.
62) Attach the vacuum line to the reservoir outlet and switch on the vacuum system.

*Tissue rehydration and nuclear staining*
Timing: 25 minutes

63) Ensure the waste line is connected to an empty container and follow institute recommendations for waste disposal.
64) Prepare the $LiBH_4$ now and make a 0.5 mg/ml solution in water. Pass the dissolved solution through a 0.22 µm syringe filter. Note the time and use within 4 hours.
65) Switch on the electric heater set to 37°C.
66) Launch the ARIA program.
    Critical: Following system priming, the imaging chamber will start to fill. Check there are no leaks. If air is not completely removed from the chamber during filling, loosen the stage insert and gently tilt so air bubbles rise to the outlet channel.

*Final setup with triggering of the THUNDER microscope*
Timing: 10 minutes

67) Wait for the Fluigent software to reach the programmed pause step. By this point the imaging chamber will be filled, the tissue will be rehydrated, and nuclear staining with Hoechst 33342 will be complete.
68) In the LAS X software, click the “Live” button to visualize the Hoechst staining and bring the center of the tissue into sharp focus.
69) Click on the “Show Highspeed Autofocus Panel” then from the Focus-system dropdown choose “Adaptive Focus Control”. Check “AFC on/off” box so it is active. Choose ‘Continuous mode’ and AFC mode ‘Quality’.
    Critical: The green light should appear and hold position remain stable. If this fails, adjust the focus up and down. Intermittent flickering is acceptable.

70) Set up THUNDER deconvolution. Recommended parameters are to use the ICC setting with RI set to 1.33. Strength is set to 98% and Feature size is best set to 2000 nm for membrane markers or 3000 nm for nuclear markers in our experience.
71) Switch on the 'Timecourse' tab and set the number of cycles to match that of the microfluidics program.
72) Enter Navigator mode, map the tissue using the spiral function and create the region of interest for acquisition by tiling.
73) While in Navigator select the triggering module.
74) With the 'ARIA to Leica' dropdown selected, check "Trigger linked to acquisition" then the check box next to tile number in the Regions Box. In the Channels tab select 'First Channel only', in the Timelapse tab select "Every cycle" and in the Stage tab select "At positions 1"
75) Switch to the 'Leica to ARIA' dropdown and check "Use in experiment." Set number of pulses to 1 and to a 100 ms duration. Check the box next to tile number in the Regions box then set the dropdown from Trigger to "After acquisition." In the Channels tab, type the total number of channels being acquired in the "At Channel Number" box. Select "Every Cycle" in the Timelapse tab and ensure the final tile number is shown under the Stage tab under "At positions."
76) Click the Start button while in Navigator mode. The software should initialize and move to the first tile position then pause and wait. Now click the popup box to continue the program in the Fluigent software.

    Critical: Make sure to check the "hold position" button again in the AFC menu if the z-position is altered at any point.

    Critical: Bubbles form on the tissue and rise to the top of the imaging chamber during fluorophore inactivation with $LiBH_4$, potentially impeding solution delivery to the tissue. We recommend manually removing air bubbles every 4 cycles. This can be done by placing a 31G tuberculin syringe in the secondary inlet port (See Figure 2b) and aspirating while PBS is injected into the imaging chamber.

    Critical: Replace the $LiBH_4$ after 4 hours of use as its activity diminishes with time. A manual pause can be programmed into the Fluigent software. The ideal workflow is to start the automated IBEX protocol early in the workday. This allows for the replacement of $LiBH_4$ and bubble removal to be completed before leaving at the end of the day, allowing the system to complete overnight.
77) When the cycle is complete the file can be saved and processed with Steps 78-90. Thoroughly clean the RC-21B chamber after each use to remove vacuum grease. Immerse in distilled water to dissolve any precipitated PBS crystals.

*Image post-processing*

Timing: 45 minutes

78) We use commercially available Imaris software (Imaris x64 9.5.0) for processing raw images. Use the Imaris File Converter x64 9.5.0 software to transform the Leica confocal microscope output .lif files into .ims files.
79) If using the automated IBEX protocol, iterative images are included as different timepoints of the same image, e.g., timepoint 1 corresponds to cycle 1 image, timepoint 2 corresponds to cycle 2 image, and so on. To obtain individual images (antibody panel 1 from cycle 1, antibody panel 2 from cycle 2) for downstream registration, go to Edit>Crop Time and select from "1" to "1" for timepoint 1 and save using a name with details related to its acquisition, e.g. human_liver_panel1 or human_liver_cycle1. Open the Imaris file with all the timepoints and repeat these steps for each cycle using from "2" to "2" to obtain the timepoint 2 image and so on.
80) For each imaging round, edit the file in Imaris and apply any desired processing steps such as image smoothing, channel naming, and channel pseudo-coloring. Channel thresholding and background subtraction can be performed using the Image Processing tools in Imaris to eliminate signal from autofluorescence or left-over signal from incomplete fluorophore inactivation.

81) The SimpleITK registration software requires use of a consistent naming strategy. Channels should be uniformly named as prefix-separator character-postfix. Example: “Panel1 CD3_AF594” with space as a separator character for the channel where an anti-CD3 Alexa Fluor 594 antibody was used in the first panel. See help file provided with extension and [https://niaid.github.io/imaris_extensions/XTRegisterSameChannel.html] for further details.
82) Perform image registration using Simple ITK for either manual (3D with z-stack) or automated (2D single z-slice) IBEX.
83) Launch the software. Click Imaris Extensions > Simple ITK > Affine Registration of z-stack using common channel.
84) One .ims file for each panel of the entire multiplex experiment should be uploaded using the Browse button next to File names. Enter the Channel name prefix separator character. If using the example above, this will be a single space. Click next.
85) Select the registration channel. This will be the common named channel used as the fiducial marker, e.g., Hoechst. Select the ‘Fixed Image’, usually the image file from cycle 1, to which other images will be registered. Select the desired directory under ‘Output file’. The ‘Start registration at resolution’ option is usually set to the maximal resolution to give the best registration possible. Reduce this if computational power is limited. Click ‘Register’ to begin.
86) After successful completion of the first step, the ‘Resample and Save Combined Image’ button will become available. Click and wait for the progress bar to complete.
87) There is a final option to generate a correlation matrix as a PDF for quantitative assessment of the registration. Check the channel box and then click ‘Compute Correlations Before and After Registration’.
88) Open the output file in Imaris. The registered image of all panels with channel information should be present and is now ready for further analysis or publication.
89) Be patient when opening the registered file for the first time. It can take 20-30 minutes to view a typical 6 cycle 20 GB dataset.
90) Immediately save the file after initial opening to speed up subsequent viewing.

**Box 1 – Tissue Specific Dissection, Grossing, and Orientation**

Individual tissues must be prepared differently to ensure correct orientation and preservation during sectioning. For the tissues highlighted in Figure 1, use the following approaches.

**Procedure**

1. Tissue Dissection and Grossing
    a. Bowel: First, pour 4% molten agarose (wt/vol) dissolved in PBS into 6 well plates, covering each well with 1-1.5 ml of agarose. Allow to harden for 20 minutes. Open the bowel—opposite the mesentery or a given lesion of interest—along its entire length using scissors. Once open, rinse any residual fecal material from the mucosal surface using a stream of PBS. Cut thin strips from the opened bowel, spanning from the mucosa to the serosa, as shown in Figure 1. Pin the bowel flat onto the hardened agarose surface using 30G needles and immediately proceed to tissue processing.
    b. Skin: These samples often curl and are difficult to orient. Shape into 1 cm x 1 cm x 3-5 mm squares and immediately proceed to tissue processing.
    c. Lymph nodes: Carefully remove excess perinodal fat with fine forceps. Small gastric lymph nodes can be kept whole, but large lymph nodes should be cut at 2-3 mm intervals perpendicular to the longest plane.
    d. Wedge resections (liver, kidney, spleen): Cut specimens into thin sections (less than 3-5 mm in thickness).
    e. Other tissues: Consult with a pathologist and the emerging Common Coordinate Framework[36] to enable relating the processed image data to tissue anatomy. Use the suggestions above for sample preparation to maintain physical integrity and effective, rapid fixation.
2. Tissue Orientation in OCT medium
    a. Bowel: Use a dissecting microscope and light source to aid in orienting the tissue for a longitudinal cross section.
    b. Skin: Position appropriately for a complete cross section perpendicular to the surface of the tissue.
    c. Lymph nodes: Orient to obtain sections along the sagittal plane.
    d. Wedge resections (liver, kidney, spleen): Identify the smooth capsule and position the sample to allow sections perpendicular to this surface.

**Box 2 – Channel Dye Separation**

Dye separation is an approach to separate fluorochromes with overlapping fluorescence spectra in multiplexed experiments, so the signals can be correctly ascribed to the markers they originate from. We recommend using the Channel Dye Separation module of the Leica LAS X software. If Leica LAS X software is not available, use an open-source alternative for spectral unmixing, e.g. ImageJ plugins[52,53] (https://www.youtube.com/watch?v=W90qs0J29v8).

**Procedure**

1. Begin by creating single color controls using reference antibodies conjugated to the same fluorophores present in IBEX tissue panels, e.g., Hoechst, CD20 AF488, CD21 AF532, CD31 PE, CD3 iF594, CD163 AF647, Ki-67 AF700).
2. Acquire single labeled tissue sections (e.g., CD20 AF488 or CD31 PE) using the same acquisition parameters employed for IBEX imaging.
3. Following acquisition of single-color control images, perform Channel Dye Separation in the LAS X Dye Separation module by selecting reference regions corresponding to positive signal for each channel (e.g., 7 channels in the above example).
4. Apply the unmixing "matrix" to all images using the Automatic Dye Separation algorithm of the Leica LAS X software. This tool uses a cluster-based analysis algorithm to vectorize the gray levels within the image and separate them into clusters for removal of crosstalk.

## Troubleshooting

| Steps | Problem | Possible Reason | Solution |
|---|---|---|---|
| 8-11, 39-40 | Tissue lifts during IBEX imaging | Insufficient chrome alum gelatin coating | Repeat with sufficient volume and coverage |
| | | Tissue not sectioned completely flat onto the imaging substrate | Repeat, ensuring the tissue does not roll or fold during sectioning |
| | | Chrome alum gelatin adhesive has expired | Check expiration date and replace with new bottle |
| | | Prepared tissue sections are too old | Image samples within 1-2 days for optimal results |
| 16-18, 48 | No or poor antibody immunolabeling | Unsuitable antibody clone or incompatible antigen retrieval conditions | Validate all antibodies first using serial, conventional imaging approaches |
| | | Insufficient concentration of antibody | Increase the µg/ml of antibody employed |
| | | Epitope is potentially sensitive to $LiBH_4$ or sterically hindered by antibodies present in earlier cycles | Move antibody to an earlier cycle or substitute with an alternative clone |
| 3-7 | High tissue autofluorescence | Tissue source and sample format high in autofluorescence, e.g. FFPE or liver tissue | Preserve tissue samples according to fixed frozen method described here. For FFPE tissues, carefully design panels with fluorophores with high quantum yields and utilize microscopes with narrow excitation/emission. See Table S5 in Hickey et al., for additional strategies[19]. |
| 50-52 | ARIA fails to prime or achieve set flow rate | Tubes or connections may be loose allowing air entry | Tighten all tubes and connection points |
| 66-67 | RC-21B imaging chamber leaks fluid | Inadequate vacuum grease | Make sure the vacuum grease forms a continuous ring without gaps |
| | | Outflow blockage | Clean the outflow port and ensure no vacuum grease is occluding it |
| 66 | Excessive air in the imaging chamber | Uneven filling during tissue rehydration | Gently loosen the chamber from the stage insert and tilt so the air shifts to the perfusion outlet port |
| 76 | Excessive air in the imaging chamber | Bubble formation during $LiBH_4$ treatment | Periodically aspirate air as needed using a tuberculin syringe via the secondary inlet port |
| 16-18, 48 | Non-specific immunolabeling or fluorescent debris collecting on the tissue | Precipitated or unbound fluorophore aggregates | Use an alternative antibody from a different vendor, always spin down antibodies with a centrifuge for 30 seconds before preparing panels, and wash tissue more extensively after labeling |
| 20, 57 | Loss of image in regions or image distortion. | Insufficient immersion oil | Apply ample amounts of immersion oil to the coverslip and microscope objective |
| 26-31, 64, 76 | Incomplete fluorophore inactivation | $LiBH_4$ was no longer active | Always use the solution within 4 hours |
| 26-31, 48, 76 | Incomplete fluorophore inactivation resulting in left-over signal in subsequent cycle | High antibody concentration for an abundant marker | Use properly titrated antibodies, place dim markers in earlier cycles, increase duration of fluorophore inactivation to 20 minutes for secondary antibodies used in automated IBEX protocol, and, if needed, threshold dim signal using image processing software post-acquisition. |
| 25, 55 | Spectral overlap of fluorescent signals between channels | Dye Separation is required | Use Dye Separation in LAS X software or equivalent software with appropriate control slides (Box 2). Use narrow bandpass filters where available |
| 78-90 | Registration software failure | 2D vs 3D settings incorrect | Under Advanced Settings check that the correct options are selected to match the dataset |

## Timing

- Steps 1-2, tissue grossing: 2-3 hours
- Steps 3-7, tissue processing: 2 days
- Steps 8-38 Manual IBEX option: Total ~28.5 hours
    - Steps 8-11, manual IBEX sample preparation: 2.5 hours
    - Steps 12-38, manual IBEX iterative imaging: interventions and wait time (~10 hours), total experimental time (26 hours) for a 24-plex dataset with the following properties: 3 $mm^2$, 4 cycle, 8 µm z-stack
- Steps 39-77 Automated IBEX option: Total ~19 hours
    - Steps 39-40, automated IBEX sample preparation: 2-3.5 hours
    - Steps 41-62, automated IBEX chamber assembly and solution preparation: 1 hour

- Steps 63-77, automated IBEX iterative imaging: interventions and wait time (40 minutes), total experimental time (14 hours) for a 24-plex dataset with the following properties: 3 $mm^2$, 6 cycle, 1 z-slice
- Steps 78-90, image post-processing: 45 minutes

**Anticipated results**

Both IBEX methods generate high quality, multiplexed imaging data with the following attributes based on the acquisition settings described here: i) image resolution of 0.284 (manual) or 0.160 µm (automated) in x-y, ii) 8-bit (manual) or 16-bit (automated) dynamic range, iii) tiled regions of interest >9 $mm^2$, and iv) total time for an equivalent 24-plex dataset of 26 (manual, 4 cycle, 8 µm z-stack) or 14 hours (automated, 6 cycle, single z-slice). These properties are distinguishing, as other imaging modalities may require 27 hours to acquire a 1 $mm^2$ area at a resolution of 0.260 µm[16]. We provide here representative datasets obtained using the manual IBEX method for a wide range of human tissues including the mesenteric lymph node (38-plex, 9 cycles), spleen (25-plex, 4 cycles), and liver (22-plex, 4 cycles) (Figure 3). Using the automated IBEX protocol, the following human fixed frozen tissues were profiled: mesenteric lymph node (24-plex, 6 cycles), jejunum (24-plex, 6 cycles), and skin (19-plex, 5 cycles), a very challenging tissue due to its delicate structure and high background autofluorescence (Figure 4a-c). Additionally, we demonstrate an application of the automated IBEX method for imaging human kidney FFPE tissues (16-plex, 5 cycles) (Figure 4d). Importantly, the cycle and marker numbers presented here are provided as a proof-of-concept and do not represent a technical limitation (see Radtke *et al.*, for IBEX imaging exceeding 65 parameters[23]). Cell-cell alignment across imaging cycles is typical and, importantly, needed for visualizing and quantifying complex cell types *in situ* as phenotypic markers are frequently distributed over multiple cycles. IBEX-generated images are compatible with established methods for analyzing high dimensional imaging data including the open source, computational histology topography cytometry analysis toolbox (histoCAT)[54], as previously demonstrated[23].

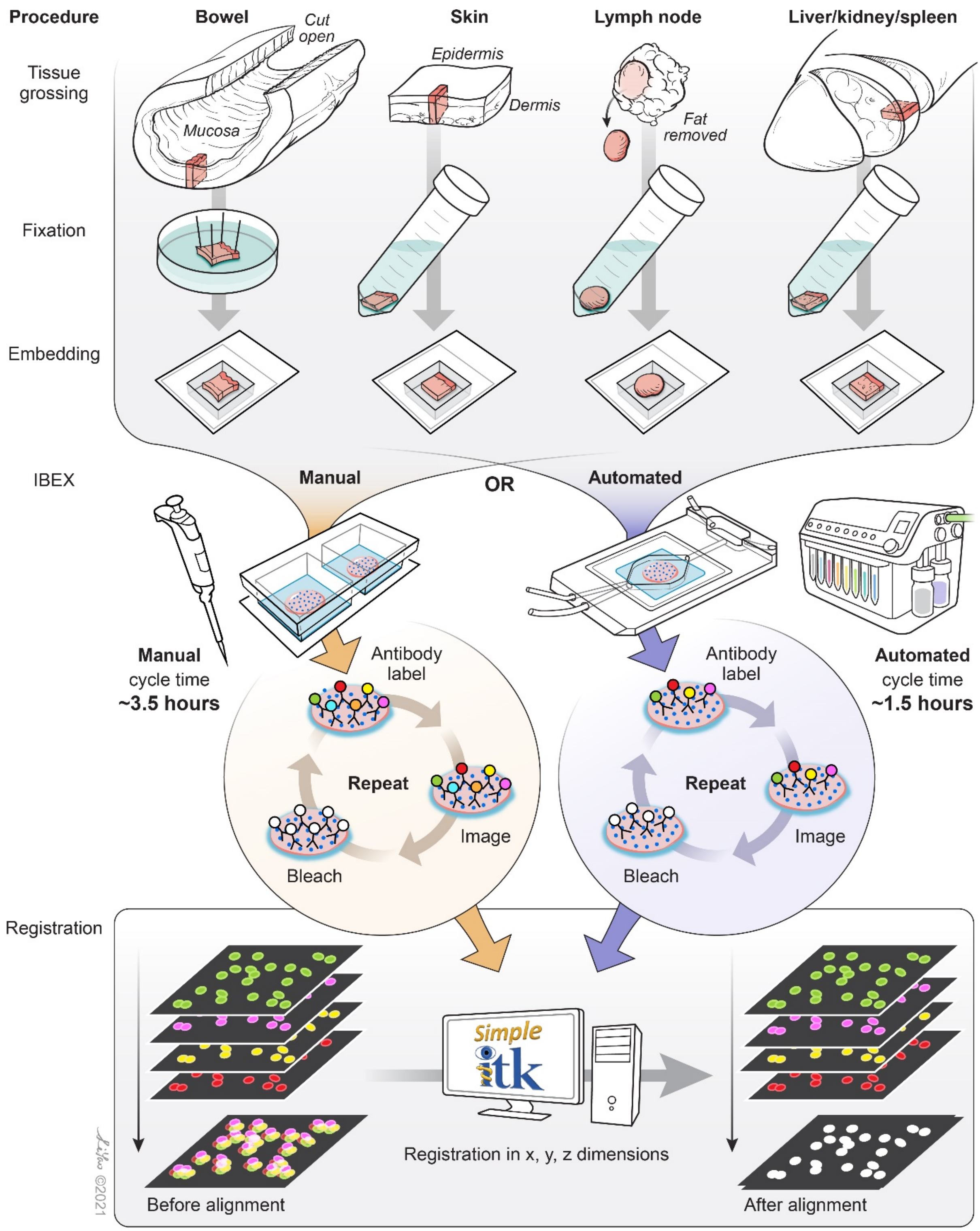

**Figure 1. Overview of tissue grossing, sample preparation, IBEX protocols, and image registration workflow.**

Specimens are first prepared using established tissue grossing protocols (Steps 1-2, Box 1, 2-3 hours). For the jejunum and skin samples, tissues are enlarged to show detail but do not reflect their actual size within the molds. Following sample preparation, tissues can be preserved as FFPE tissues (not shown) or as fixed frozen samples (Steps 3-7, 2 days). High content imaging is performed using the manual (Steps 8-38, 2-4 days) or automated (Steps 39-77, 18 hours) IBEX method, consisting of iterative cycles of antibody labeling, imaging, and fluorophore inactivation with $LiBH_4$. IBEX-generated images are then processed and aligned using SimpleITK open source software (Steps 78-90, 45 minutes).

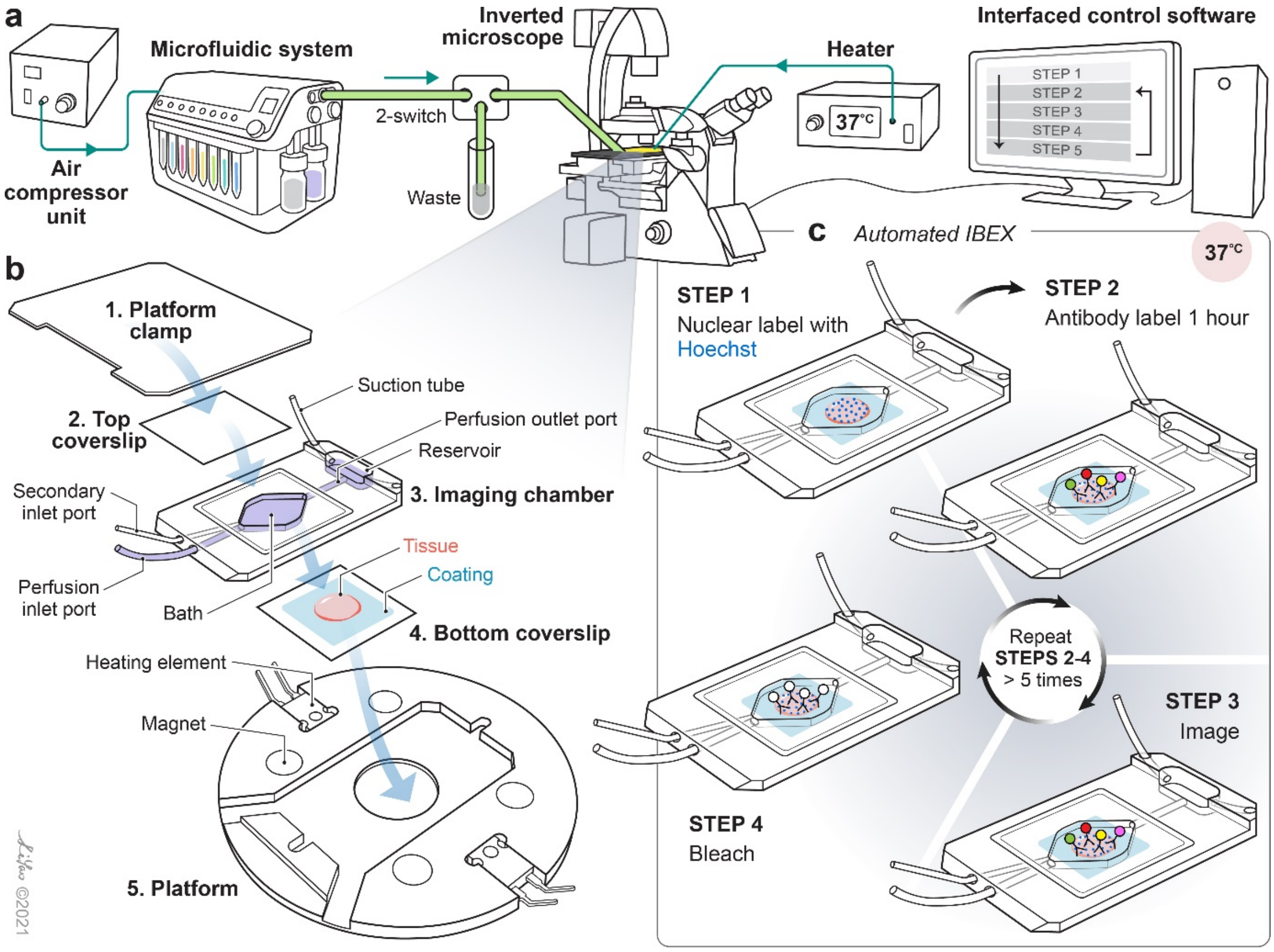

**Figure 2. Schematic overview of automated IBEX protocol.**

(***a***) The automated IBEX protocol uses a compact microfluidics system that delivers multiple solutions to an imaging chamber placed on an inverted microscope stage. Fluids are pushed through the system with an air compressor unit and delivered to the imaging chamber based on signals coordinated by the chip (2-switch). The inverted microscope and fluidics device send and receive TTL pulses using precise timings established by interfaced control software. (***b***) Samples are sectioned onto coated coverslips and assembled into a closed bath imaging chamber. Following assembly, the imaging chamber is secured to a magnetic platform and mounted onto the microscope stage. Schematic demonstrates the multiple components of the overall sample preparation from top (1) to bottom (5). Fluid (purple shading) enters the chamber via the perfusion inlet port, collects in the bath, and exits by the perfusion outlet port that is attached to a vacuum line (not pictured). (***c***) Automated IBEX consists of: 1) nuclear labeling with Hoechst, 2) antibody labeling for 1 hour at 37°C using a heated microscope stage, 3) imaging region(s) of interest, 4) bleaching with $LiBH_4$, and 5) repeating steps 2-4 until the desired number of parameters is achieved, typically 12 hours for a 6 cycle, 25-plex experiment.

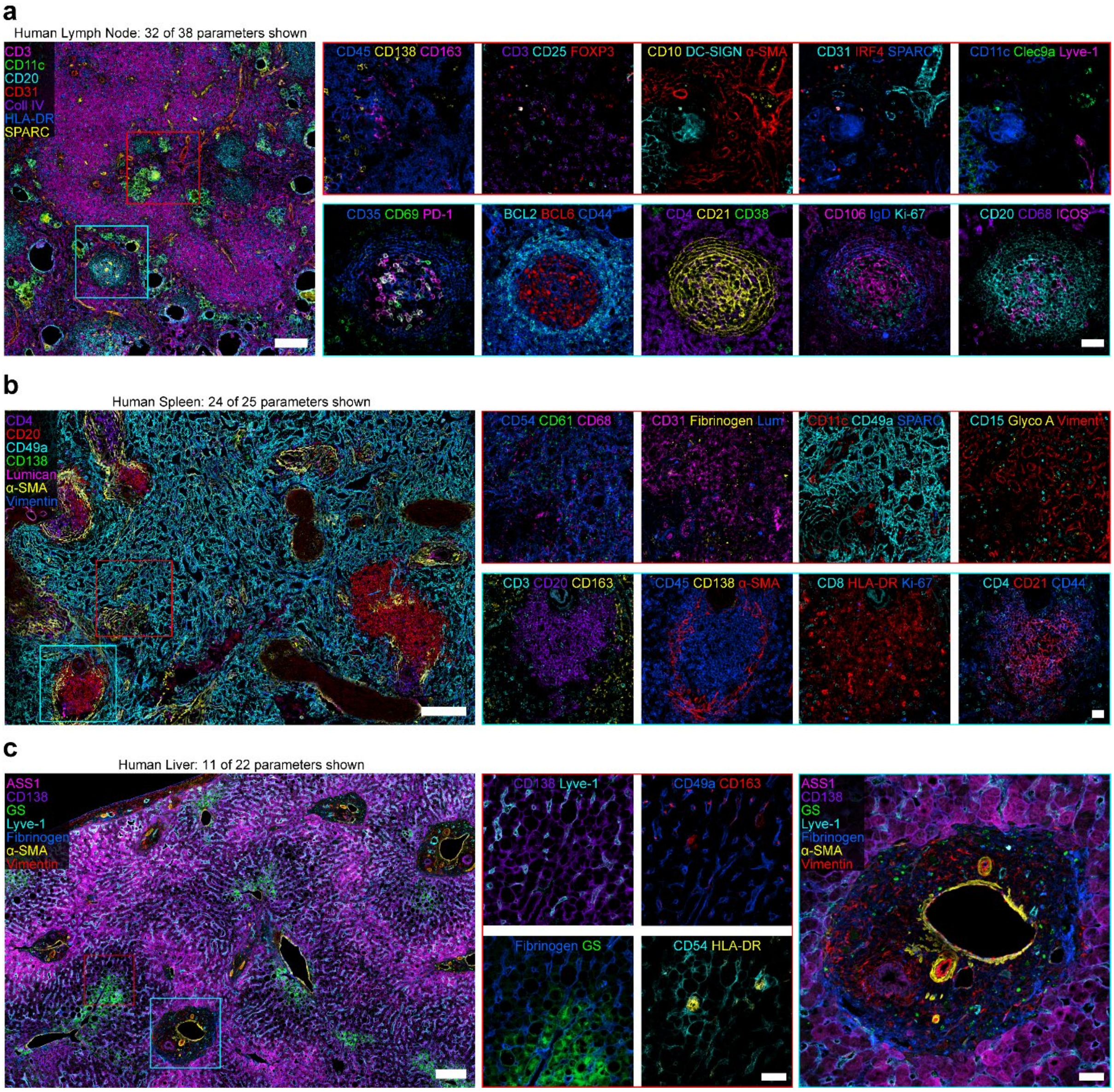

**Figure 3. Representative images of manual IBEX method in human tissues.**

***(a)*** Confocal images from a human mesenteric LN (9 cycles, 32 of 38 parameters shown). Scale bar is 200 μm (left), 25 μm (insets). ***(b)*** Confocal images from human spleen (4 cycles, 24 of 25 parameters shown). Scale bar is 200 μm (left), 25 μm (insets). Glycophorin A (Glyco A), Lumican (Lum), and Vimentin (Viment). (**c**) Confocal images from human liver (4 cycles, 11 of 22 parameters shown). Scale bar is 200 μm (left), 50 μm (insets). Glutamine synthetase (GS).

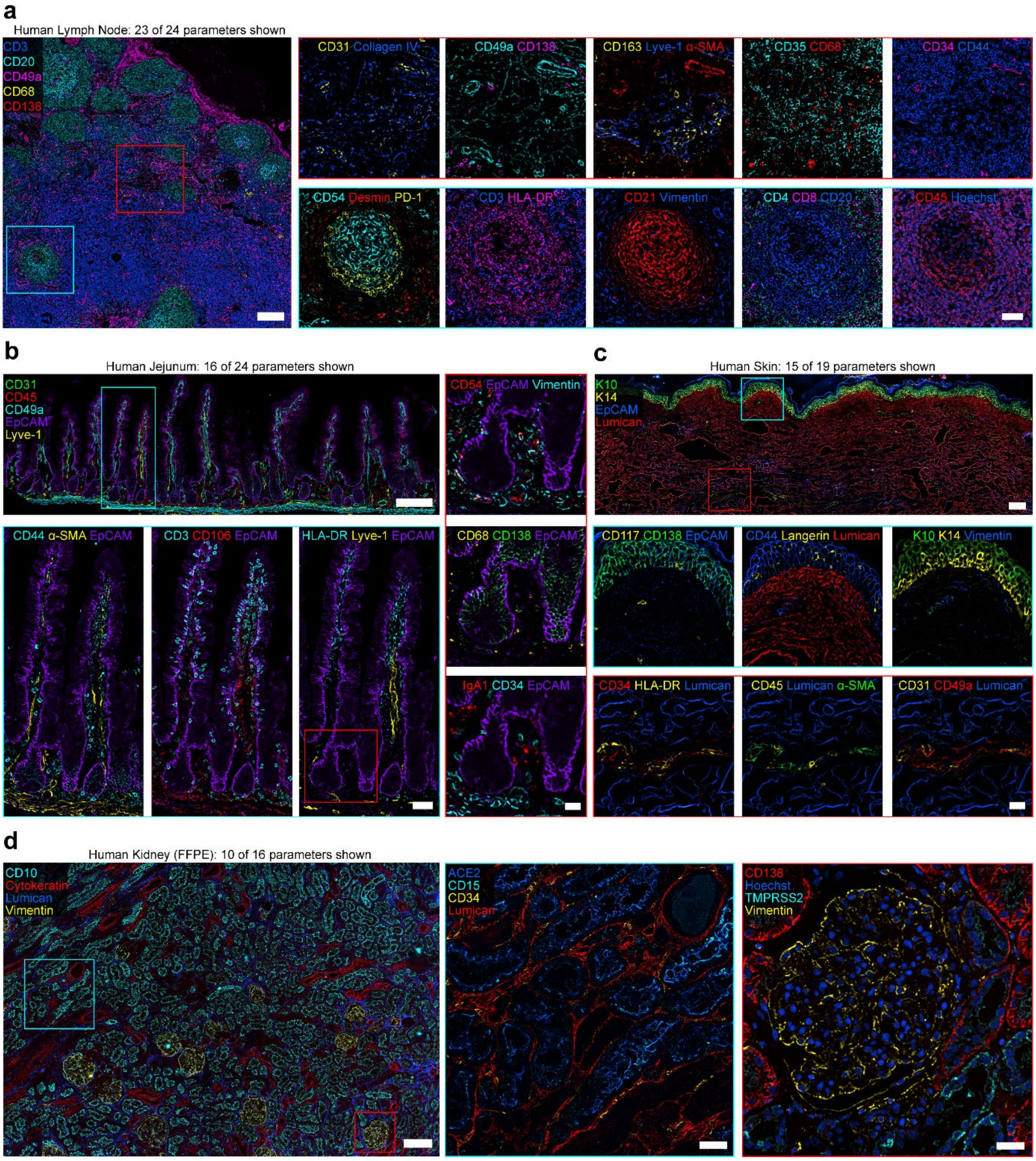

**Figure 4. Representative images of automated IBEX method in human tissues.**
***(a)*** Images from a human mesenteric LN (6 cycles, 23 of 24 parameters shown). Scale bar is 200 µm (left), 50 µm (insets). ***(b)*** Images from human jejunum (6 cycles, 16 of 24 parameters shown). Scale bar is 200 µm (left), 50 µm (cyan box), 25 µm (red box). (**c**) Images from human skin (5 cycles, 15 of 19 parameters shown). Scale bar is 200 µm (left), 25 µm (insets). Keratin 10 (K10), Keratin 14 (K14). (***d***) Images from FFPE kidney section (5 cycles, 10 of 16 parameters shown). Scale bar is 200 µm (left), 50 µm (cyan box), 25 µm (red box).

### Author contributions statement

A.J.R, C.J.C., and R.N.G. wrote the manuscript. A.J.R., C.J.C., H.I., and R.T.B. designed and executed the experiments. Z.R.Y. and B.L. developed image analysis software. L.Y. designed figures 1 and 2 and A.J.R. designed figures 3 and 4. J.M. integrated the Leica microscope with the fluidics device. A.G., J.K., E.S., N.T., J.C., D.J., J.D., and J.M.H. provided technical insight, reagents, and tissues. All authors offered their guidance for the development and optimization of the workflows.

### Acknowledgements

This research was supported by the Intramural Research Program of the NIH, NIAID and NCI. This research was also partially supported by a Research Collaboration Agreement (RCA) between NIAID and BioLegend, Inc. (RCA# 2020-0333) and the Chan Zuckerberg Initiative Human Cell Atlas Thymus Seed Network. C.J.C is supported as a UK-US Fulbright Scholar and Fight for Sight Research Scholar. Z.Y. and B.C.L. are supported by the BCBB Support Services Contract HHSN316201300006W/HHSN27200002 to Medical Science & Computing, LLC. D.J. is supported by the grant of the European Research Council (ERC); European Consolidator Grant, XHale (Reference #771883). We would like to thank Robert Pelletier and Mark Aruda from Fluigent for their sterling assistance with the ARIA fluidics device. We are extremely grateful for the technical support provided by George Portugal, Emily Cox, and Ed Buck from Harvard Apparatus. We thank Dr. Stefania Pittaluga for her assistance with tissue grossing and orientation. We are appreciative of Drs. Giorgio Cattoretti and Maddalena Bolognesi for sharing their insights on fluorophore inactivation with sodium borohydride.

### Competing interests

Joshua Croteau is an employee of Biolegend, Inc. and James Marr is an employee of Leica Microsystems, Inc.

**Supplementary Information**

Extended Data Figure 1. Critical steps in the manual IBEX protocol.
Extended Data Figure 2. Equipment and assembly of imaging chamber for automated IBEX protocol.
Table S1. Time and method used to bleach fluorescently conjugated antibodies and dyes.
Table S2. IBEX panels for human organs.
Table S3. Recommendations for optimal antibody and fluorophore pairing for IBEX panel design.
Table S4. Estimated costs for IBEX implementation.

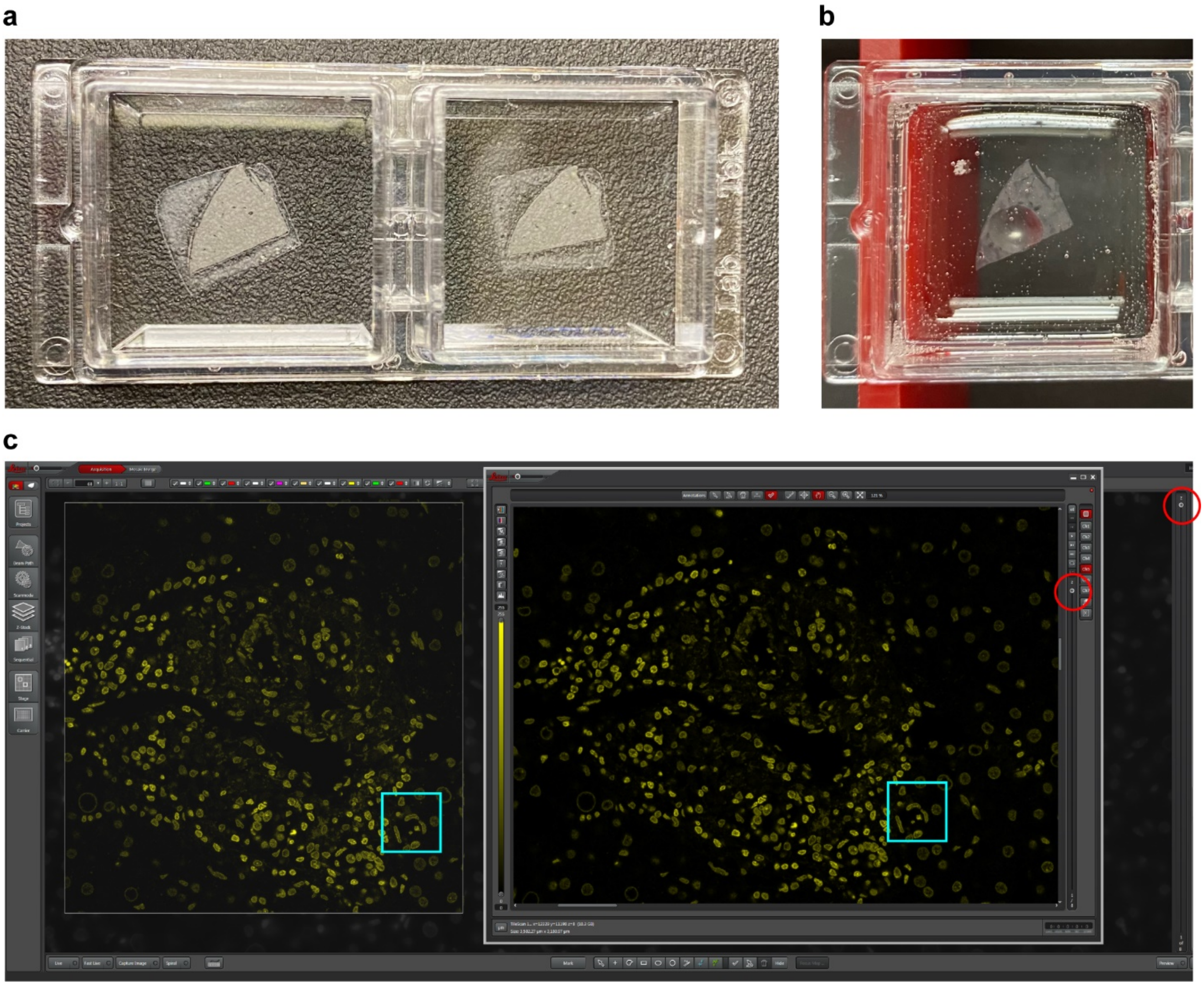

**Extended Data Figure 1. Critical steps in the manual IBEX protocol.**

(***a***) Photo depicting central placement of tissue within a 2-well chambered coverglass. Glass surface is coated with chrome gelatin alum (invisible when fully dry). (***b***) Picture of small bubbles that form during successful $LiBH_4$ treatment. (***c***) Visual instructions on how to match unique nuclear shapes (Hoechst in yellow, blue box) across the imaging volumes. The left image corresponds to a live image in the Leica LAS X Navigator software. The right image corresponds to the image captured from the previous IBEX cycle. Red circles indicate that the described alignment procedure is being done at the first z slice ('Begin') of the z-stack.

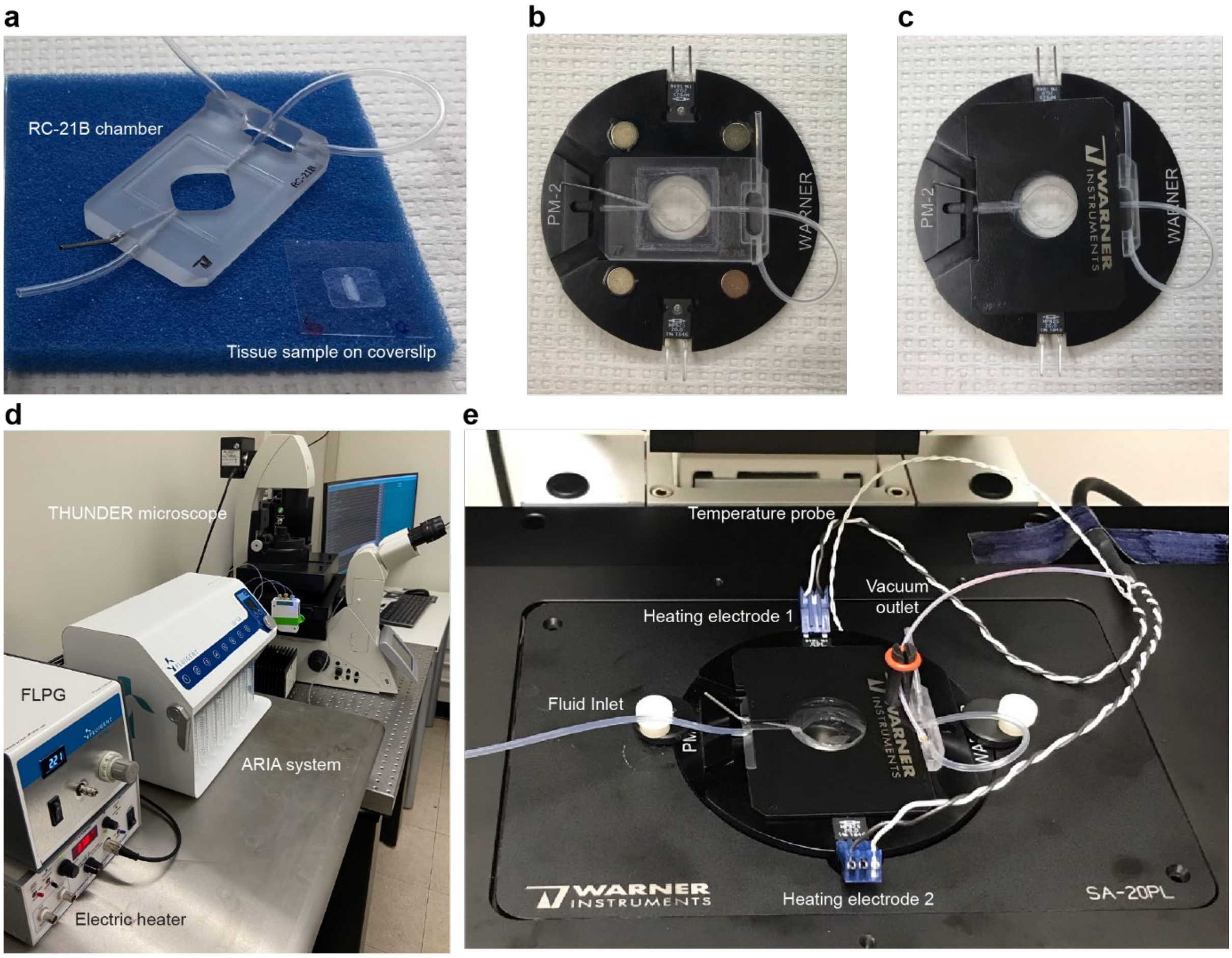

**Extended Data Figure 2. Equipment and assembly of imaging chamber for automated IBEX protocol.**

(***a***) Tissues are sectioned onto coated 22x22 mm square coverslips and assembled into the RC-21B Large Closed Bath Imaging Chamber. Top view of imaging chamber placed into PM-2 Platform for Series 20 chambers without (***b***) and with (***c***) magnetic platform clamp. (***d***) Equipment footprint of automated IBEX set up; Fluigent Low Pressure Generator (FLPG). (***e***) Complete assembly of PM-2 Platform with RC-21B chamber onto SA20-PL Series 20 stage adapter. Fluid inlet and vacuum outlet highlight the fluid path. Heating electrodes are attached to the top and bottom of the platform using metal prongs that must be bent by user to allow placement into the stage. Temperature probe is inserted into small hole at top of platform to maintain 37°C for the duration of the protocol.

# Tables S1-S4

## Table S1. Time and method used to bleach fluorescently conjugated antibodies and dyes.

| | | | | | | | | Time to Bleach (Minutes) | |
|---|---|---|---|---|---|---|---|---|---|
| **Antibody** | **Clone** | **RRID** | **Vendor** | **Cat No.** | **Species reactivity** | **Isotype** | **Dilution** | $LiBH_4$ | $LiBH_4$ + Light |
| β-3 Tubulin BV421 | AA10 | AB_2632699 | BioLegend | 657412 | Human, Mouse | Mouse IgG2a, κ | 1:200 | - | <15 |
| β-3 Tubulin AF488 | TUJ1 | AB_2564757 | BioLegend | 801203 | Human, Mouse | Mouse IgG2a, κ | 1:50 | <15 | - |
| β-3 Tubulin AF532 | AA10 | NA | BioLegend | Custom | Human, Mouse | Mouse IgG2a, κ | 1:100 | <15 | - |
| ACE2 | - | AB_355722 | R&D | AF933 | Human | Goat IgG | 1:50 | - | - |
| B220 BV421 | RA3-6B2 | AB_2562905 | BioLegend | 103251 | Mouse | Rat IgG2a, κ | 1:400 | - | <15 |
| B220 BV510 | RA3-6B2 | AB_2650679 | BioLegend | 103248 | Mouse | Rat IgG2a, κ | 1:300 | - | <15 |
| B220 AF488 | RA3-6B2 | AB_396781 | BD Biosciences | 557669 | Mouse | Rat IgG2a, κ | 1:400 | <15 | - |
| B220 AF532 | RA3-6B2 | AB_11218492 | Thermo | 58-0452-82 | Mouse | Rat IgG2a, κ | 1:50 | <15 | - |
| B220 PE | RA3-6B2 | AB_394620 | BD Biosciences | 553090 | Mouse | Rat IgG2a, κ | 1:400 | <15 | - |
| B220 eF570 | RA3-6B2 | AB_2573598 | Thermo | 41-0452-80 | Mouse | Rat IgG2a, κ | 1:200 | <15 | - |
| B220 eF615 | RA3-6B2 | AB_10852702 | Thermo | 42-0452-82 | Mouse | Rat IgG2a, κ | 1:200 | >120 | - |
| B220 AF647 | RA3-6B2 | AB_389330 | BioLegend | 103226 | Mouse | Rat IgG2a, κ | 1:400 | <15 | - |
| B220 AF700 | RA3-6B2 | AB_493717 | BioLegend | 103232 | Mouse | Rat IgG2a, κ | 1:50 | <15 | - |
| BCL2 AF647 | 100 | AB_2563279 | BioLegend | 658705 | Human | Mouse $IgG_1$ | 1:25 | <15 | - |
| Bcl6 AF647 | K112-91 | AB_10898007 | BD Biosciences | 561525 | Human, Mouse | Mouse $IgG_1$, κ | 1:25 | <15 | - |
| Cathepsin L | 204101 | AB_2087829 | R&D | MAB-9521 | Human | Rat IgG2a | 1:50 | - | - |
| CD1c PE | L161 | AB_1089000 | BioLegend | 331506 | Human | Mouse $IgG_1$, κ | 1:50 | <15 | - |
| CD1d PE | 1B1 | AB_2073521 | BD Biosciences | 553846 | Mouse | Rat IgG2b, κ | 1:100 | <15 | - |
| CD3 BV421 | 17A2 | AB_2562553 | BioLegend | 100228 | Mouse | Rat IgG2b, κ | 1:400 | - | <15 |
| CD3 BV510 | 17A2 | AB_2562555 | BioLegend | 100234 | Mouse | Rat IgG2b, κ | 1:50 | - | <15 |
| CD3 AF488 | 17A2 | AB_389301 | BioLegend | 100210 | Mouse | Rat IgG2b, κ | 1:200 | <15 | - |
| CD3 AF532 | 17A2 | AB_11220474 | Thermo | 58-0032-80 | Mouse | Rat IgG2b, κ | 1:50 | <15 | - |
| CD3 PE | 17A2 | AB_312662 | BioLegend | 100205 | Mouse | Rat IgG2b, κ | 1:200 | <15 | - |
| CD3 AF594 | 17A2 | AB_2563427 | BioLegend | 100240 | Mouse | Rat IgG2b, κ | 1:400 | >120 | - |
| CD3 AF647 | 17A2 | AB_396912 | BD Biosciences | 557869 | Mouse | Rat IgG2b, κ | 1:400 | <15 | - |
| CD3 AF532 | UCHT1 | AB_11218675 | Thermo | 58-0038-42 | Human | Mouse $IgG_1$, κ | 1:50 | <15 | - |
| CD3 AF594 | UCHT1 | AB_2563236 | BioLegend | 300446 | Human | Mouse $IgG_1$ | 1:200 | >120 | - |
| CD3 iF594 | UCHT1 | NA | Caprico Biotechnologies | 1053134 | Human | Mouse $IgG_1$ | 1:50 | <15 | - |
| CD4 BV421 | GK1.5 | AB_2562557 | BioLegend | 100443 | Mouse | Rat IgG2b, κ | 1:200 | - | <15 |
| CD4 BV510 | GK1.5 | AB_2564587 | BioLegend | 100449 | Mouse | Rat IgG2b, κ | 1:50 | - | <15 |
| CD4 PE | GK1.5 | AB_395014 | BD Biosciences | 553730 | Mouse | Rat IgG2b, κ | 1:200 | <15 | - |
| CD4 AF594 | GK1.5 | AB_2563182 | BioLegend | 100446 | Mouse | Rat IgG2b, κ | 1:200 | >120 | - |
| CD4 AF488 | RM4-5 | AB_396779 | BD Biosciences | 557667 | Mouse | Rat IgG2a, κ | 1:100 | <15 | - |
| CD4 AF532 | RM4-5 | AB_11219484 | Thermo | 58-0042-80 | Mouse | Rat IgG2a, κ | 1:50 | <15 | - |
| CD4 eF570 | RM4-5 | AB_2573595 | Thermo | 41-0042-82 | Mouse | Rat IgG2a, κ | 1:100 | <15 | - |
| CD4 AF532 | RPA-T4 | AB_2802361 | Thermo | 58-0049-42 | Human | Mouse $IgG_1$, κ | 1:25 | <15 | - |
| CD4 AF647 | RPA-T4 | AB_389333 | BioLegend | 300520 | Human | Mouse $IgG_1$, κ | 1:100 | <15 | - |
| CD4 AF700 | RPA-T4 | AB_493743 | BioLegend | 300526 | Human | Mouse $IgG_1$, κ | 1:25 | <15 | - |
| CD8 BV421 | 53-6.7 | AB_11204079 | BioLegend | 100738 | Mouse | Rat IgG2a, κ | 1:200 | - | <15 |
| CD8 BV510 | 53-6.7 | AB_2563057 | BioLegend | 100752 | Mouse | Rat IgG2a, κ | 1:200 | - | <15 |
| CD8 AF488 | 53-6.7 | AB_389304 | BioLegend | 100723 | Mouse | Rat IgG2a, κ | 1:200 | <15 | - |
| CD8 PE | 53-6.7 | AB_394570 | BD Biosciences | 553032 | Mouse | Rat IgG2a, κ | 1:400 | <15 | - |
| CD8 AF594 | 53-6.7 | AB_2563237 | BioLegend | 100758 | Mouse | Rat IgG2a, κ | 1:200 | >120 | - |
| CD8 AF647 | 53-6.7 | AB_389326 | BioLegend | 100724 | Mouse | Rat IgG2a, κ | 1:200 | <15 | - |
| CD8 AF488 | SK1 | AB_10549301 | BioLegend | 344716 | Human | Mouse IgG$_1$, κ | 1:50 | <15 | - |
| CD10 PE | FR4D11 | NA | Caprico Biotechnologies | 103926 | Human | Mouse IgG$_1$, κ | 1:50 | <15 | - |
| CD10 | - | AB_354652 | R&D | AF1182 | Human | Goat IgG | 1:200 | - | - |
| CD11b FITC | 5C6 | AB_2538034 | Thermo | MA5-16529 | Mouse | Rat IgG2b | 1:100 | <30 | - |
| CD11b PE | 5C6 | AB_323678 | Bio-Rad | MCA711PE | Mouse | Rat IgG2b | 1:100 | <15 | - |
| CD11b AF488 | M1/70 | AB_389305 | BioLegend | 101217 | Mouse | Rat IgG2b, κ | 1:100 | <15 | - |
| CD11c AF488 | N418 | AB_10373244 | Thermo | MCD11c20 | Mouse | Hamster IgG | 1:50 | <15 | - |
| CD11c AF594 | N418 | AB_2563323 | BioLegend | 117346 | Mouse | Hamster IgG | 1:50 | >120 | - |
| CD11c AF647 | N418 | AB_389328 | BioLegend | 117312 | Mouse | Hamster IgG | 1:100 | <15 | - |
| CD11c AF700 | B-Ly6 | AB_10612006 | BD Biosciences | 561352 | Human | Mouse IgG$_1$, κ | 1:25 | <15 | - |
| CD15 | MMA | AB_400298 | BD Biosciences | 347420 | Human | Mouse IgM | 1:50 | - | - |
| CD20 AF488 | L26 | AB_10734358 | Thermo | 53-0202-82 | Human | Mouse IgG2b, κ | 1:200 | <15 | - |
| CD20 eF615 | L26 | AB_10853517 | Thermo | 42-0202-82 | Human | Mouse IgG2b, κ | 1:200 | >120 | - |
| CD20 eF660 | L26 | AB_11150959 | Thermo | 50-0202-82 | Human | Mouse IgG2b, κ | 1:200 | <15 | - |
| CD21 Pacific Blue | 7E9 | AB_2085159 | BioLegend | 123413 | Mouse | Rat IgG2a, κ | 1:200 | <15 | - |
| CD21 AF488 | Bu32 | NA | BioLegend | Custom | Human | Mouse IgG$_1$, κ | 1:100 | <15 | - |
| CD21 AF532 | Bu32 | NA | BioLegend | Custom | Human | Mouse IgG$_1$, κ | 1:400 | <15 | - |
| CD23 AF647 | B3B4 | AB_493479 | BioLegend | 101611 | Mouse | Rat IgG2a, κ | 1:50 | <15 | - |
| CD23 AF532 | EBVCS-5 | NA | BioLegend | Custom | Human | Mouse IgG$_1$, κ | 1:25 | <15 | - |
| CD25 AF488 | PC61.5 | AB_763472 | Thermo | 53-0251-82 | Mouse | Rat IgG1, λ | 1:50 | <15 | - |
| CD25 AF647 | M-A251 | AB_2563587 | BioLegend | 356127 | Human | Mouse IgG$_1$, κ | 1:50 | <15 | - |
| CD31 AF488 | MEC13.3 | AB_2161031 | BioLegend | 102514 | Mouse | Rat IgG2a, κ | 1:100 | <15 | - |
| CD31 PE | MEC13.3 | AB_394819 | BD Biosciences | 553373 | Mouse | Rat IgG2a, κ | 1:200 | <15 | - |
| CD31 AF594 | MEC13.3 | AB_2563319 | BioLegend | 102520 | Mouse | Rat IgG2a, κ | 1:100 | >120 | - |
| CD31 AF647 | MEC13.3 | AB_2161029 | BioLegend | 102516 | Mouse | Rat IgG2a, κ | 1:100 | <15 | - |
| CD31 PE | WM59 | AB_314332 | BioLegend | 303106 | Human | Mouse IgG$_1$, κ | 1:100 | <15 | - |
| CD31 AF700 | WM59 | AB_2566326 | BioLegend | 303133 | Human | Mouse IgG$_1$, κ | 1:25 | <15 | - |
| CD34 PE | QBEND/10 | AB_11152571 | Thermo | MA1-10205 | Human | Mouse IgG$_1$ | 1:50 | <15 | - |
| CD34 iF594 | 4H11 | NA | AAT Bioquest | 103400C0 | Human | Mouse IgG$_1$ | 1:25 | <15 | - |
| CD35 BV510 | 8C12 | AB_2739889 | BD Biosciences | 740132 | Mouse | Rat IgG2a, κ | 1:600 | - | <15 |
| CD35 PE | E11 | AB_2292231 | BioLegend | 333406 | Human | Mouse IgG$_1$, κ | 1:800 | <15 | - |
| CD38 AF700 | HIT2 | AB_2072781 | BioLegend | 303524 | Human | Mouse IgG$_1$, κ | 1:25 | <15 | - |
| CD39 PE | A1 | AB_940429 | BioLegend | 328208 | Human | Mouse IgG$_1$, κ | 1:50 | <15 | - |
| CD44 AF488 | IM7 | AB_493679 | BioLegend | 103016 | Human, Mouse | Rat IgG2b, κ | 1:100 | <15 | - |
| CD44 AF532 | IM7 | NA | BioLegend | Custom | Human, Mouse | Rat IgG2b, κ | 1:100 | <15 | - |
| CD44 AF647 | IM7 | AB_493681 | BioLegend | 103018 | Human, Mouse | Rat IgG2b, κ | 1:100 | <15 | - |
| CD44 AF700 | IM7 | AB_493713 | BioLegend | 103026 | Human, Mouse | Rat IgG2b, κ | 1:50 | <15 | - |
| CD45 BV421 | 30-F11 | AB_2562559 | BioLegend | 103134 | Mouse | Rat IgG2b, κ | 1:100 | - | <15 |
| CD45 BV510 | 30-F11 | AB_2563061 | BioLegend | 103138 | Mouse | Rat IgG2b, κ | 1:100 | - | <15 |
| CD45 AF488 | 30-F11 | AB_493531 | BioLegend | 103122 | Mouse | Rat IgG2b, κ | 1:200 | <15 | - |
| CD45 AF532 | 30-F11 | AB_11218871 | Thermo | 58-0451-82 | Mouse | Rat IgG2b, κ | 1:200 | <15 | - |
| CD45 AF647 | 30-F11 | AB_493533 | BioLegend | 103124 | Mouse | Rat IgG2b, κ | 1:400 | <15 | - |
| CD45 AF700 | 30-F11 | AB_493715 | BioLegend | 103128 | Mouse | Rat IgG2b, κ | 1:50 | <15 | - |
| CD45 AF532 | HI30 | AB_11218084 | Thermo | 58-0459-41 | Human | Mouse $IgG_1$, κ | 1:50 | <15 | - |
| CD45 AF594 | HI30 | AB_2629599 | BioLegend | 304060 | Human | Mouse $IgG_1$, κ | 1:100 | >120 | - |
| CD45 PE/iFluor594 | F10-89-4 | NA | Caprico Biotechnologies | 1016185 | Human | Mouse IgG2a, κ | 1:50 | <15 | - |
| CD45 iF594 | F10-89-4 | NA | Caprico Biotechnologies | 1016136 | Human | Mouse IgG2a, κ | 1:50 | <15 | - |
| CD49a PE | TS2/7 | AB_1236441 | BioLegend | 328304 | Human | Mouse $IgG_1$, κ | 1:50 | <15 | - |
| CD49a AF647 | TS2/7 | AB_2129242 | BioLegend | 328310 | Human | Mouse $IgG_1$, κ | 1:200 | <15 | - |
| CD54 AF647 | HA58 | AB_2715942 | BioLegend | 353114 | Human | Mouse $IgG_1$, κ | 1:50 | <15 | - |
| CD61 FITC | Y2/51 | AB_2660493 | Miltenyi | 130-098-682 | Human | Mouse $IgG_1$, κ | 1:25 | <30 | - |
| CD64 AF647 | X54-5/7.1 | AB_2566561 | BioLegend | 139322 | Mouse | Mouse $IgG_1$, κ | 1:50 | <15 | - |
| CD66b iF594 | G10F5 | NA | Caprico Biotechnologies | 1075139 | Human | Mouse IgM, κ | 1:25 | <15 | - |
| CD66b AF647 | G10F5 | AB_2563171 | BioLegend | 305110 | Human | Mouse IgM, κ | 1:25 | <15 | - |
| CD68 BV421 | FA-11 | AB_2562949 | BioLegend | 137017 | Mouse | Rat IgG2a | 1:200 | - | <15 |
| CD68 AF488 | FA-11 | AB_2074847 | BioLegend | 137011 | Mouse | Rat IgG2a | 1:50 | <15 | - |
| CD68 iF594 | KP1 | NA | Caprico Biotechnologies | 1064135 | Human | Mouse $IgG_1$, κ | 1:100 | <15 | - |
| CD68 AF647 | KP1 | AB_627158 | Santa Cruz | sc-200060 | Human | Mouse $IgG_1$, κ | 1:100 | <15 | - |
| CD69 | - | AB_416586 | R&D | AF2386 | Mouse | Goat IgG | 1:50 | - | - |
| CD69 AF647 | FN50 | AB_528871 | BioLegend | 310918 | Human | Mouse $IgG_1$, κ | 1:50 | <15 | - |
| CD94 PE | DX22 | AB_314536 | BioLegend | 305506 | Human | Mouse $IgG_1$, κ | 1:50 | <15 | - |
| CD103 AF488 | 2E7 | AB_535949 | BioLegend | 121407 | Mouse | Hamster IgG | 1:100 | <15 | - |
| CD106 AF488 | 429 | AB_493427 | BioLegend | 105710 | Mouse | Rat IgG2a, κ | 1:50 | <15 | - |
| CD106 eF660 | 429 | AB_11217676 | Thermo | 50-1061-80 | Mouse | Rat IgG2a, κ | 1:50 | <15 | - |
| CD106 PE | STA | AB_314561 | BioLegend | 305806 | Human | Mouse $IgG_1$, κ | 1:50 | <15 | - |
| CD117 BV421 | 2B8 | AB_10898120 | BioLegend | 105827 | Mouse | Rat IgG2b, κ | 1:50 | - | <15 |
| CD117 AF488 | 104D2 | AB_2566221 | BioLegend | 313234 | Human | Mouse $IgG_1$, κ | 1:50 | <15 | - |
| CD117 PE | 104D2 | AB_314983 | BioLegend | 313204 | Human | Mouse $IgG_1$, κ | 1:50 | <15 | - |
| CD138 BV421 | 281-2 | AB_11153126 | BD Biosciences | 562610 | Mouse | Rat IgG2a, κ | 1:50 | - | <15 |
| CD138 AF647 | 281-2 | AB_2566239 | BioLegend | 142526 | Mouse | Rat IgG2a, κ | 1:50 | <15 | - |
| CD138 PE | MI15 | AB_2561878 | BioLegend | 356504 | Human | Mouse $IgG_1$, κ | 1:200 | <15 | - |
| CD138 AF647 | MI15 | AB_2564250 | BioLegend | 356523 | Human | Mouse $IgG_1$, κ | 1:200 | <15 | - |
| CD163 PE | GH1/61 | AB_1134002 | BioLegend | 333606 | Human | Mouse $IgG_1$, κ | 1:300 | <15 | - |
| CD163 AF647 | GH1/61 | AB_2563474 | BioLegend | 333620 | Human | Mouse $IgG_1$, κ | 1:100 | <15 | - |
| CD166 AF488 | EPR2759 | AB_2889210 | Abcam | Ab197543 | Human | Rabbit mAb | 1:100 | <15 | - |
| CD169 FITC | 3D6.112 | AB_324246 | Bio-Rad | MCA884F | Mouse | Rat IgG2a, κ | 1:50 | <30 | - |
| CD169 PE | 3D6.112 | AB_10915697 | Biolegend | 142404 | Mouse | Rat IgG2a, κ | 1:300 | <15 | - |
| CD169 AF594 | 3D6.112 | AB_2565620 | BioLegend | 142416 | Mouse | Rat IgG2a, κ | 1:100 | >120 | - |
| CD200 BV421 | Ox-90 | AB_2739289 | BD Biosciences | 565547 | Mouse | Rat IgG2a, κ | 1:100 | - | <15 |
| CD206 BV421 | C068C2 | AB_2562232 | BioLegend | 141717 | Mouse | Rat IgG2a, κ | 1:100 | - | <15 |
| CD207 PE | eBioL31 | AB_763453 | Thermo | 12-2075-80 | Mouse | Rat IgG2a | 1:50 | <15 | - |
| Clec9a AF488 | 8F9 | NA | BioLegend | Custom | Human | Mouse IgG2a, κ | 1:25 | <15 | - |
| Collagen IV | - | AB_445160 | Abcam | ab19808 | Mouse | Rabbit IgG | 1:200 | - | - |
| Collagen IV | - | AB_305584 | Abcam | ab6586 | Human | Rabbit IgG | 1:200 | - | - |
| CXCL12 AF532 | 79018 | - | R&D | MAB350-500 (Unconjugated) | Human | Mouse $IgG_1$ | 1:25 | <15 | - |
| CXCL13 AF532 | - | AB_355613 | R&D | AF801 (Unconjugated) | Human | Goat IgG | 1:25 | <15 | - |
| CXCR6 AF647 | 221002 | NA | Novus | FAB2145R | Mouse | Rat IgG2b | 1:50 | <15 | - |
| Cytokeratin AF647 | C-11 | AB_2563652 | BioLegend | 628604 | Human | Mouse IgG$_1$, κ | 1:200 | <15 | - |
| Cytokeratin AF488 | AE1/AE3 | AB_2574301 | Thermo | 53-9003-82 | Human, Mouse | Mouse IgG$_1$ | 1:50 | <15 | - |
| Cytokeratin eF660 | AE1/AE3 | AB_2574301 | Thermo | 50-9003-82 | Human, Mouse | Mouse IgG$_1$ | 1:100 | <15 | - |
| Cytokeratin 7 AF488 | | AB_2728458 | BioLegend | 601605 | Human | Rat IgG2a, κ | 1:50 | <15 | - |
| Cytokeratin 10 AF488 | DE-K10 | NA | Novus | NBP2-54402AF488 | Human | Mouse IgG$_1$, κ | 1:200 | <15 | - |
| Cytokeratin 14 | Poly9060 | AB_2616962 | BioLegend | 906004 | Human, Mouse | Chicken IgY | 1:50 | - | - |
| DCAMKL1 | - | AB_873538 | Abcam | Ab37994 | Human, Mouse | Rabbit IgG | 1:50 | - | - |
| DC-SIGN AF647 | 9E9A8 | AB_1186092 | BioLegend | 330112 | Human | Mouse IgG2a, κ | 1:50 | <15 | - |
| DEC205 AF647 | NLDC-145 | AB_2137655 | BioLegend | 138204 | Mouse | Rat IgG2a, κ | 1:50 | <15 | - |
| Desmin | - | AB_301744 | Abcam | Ab15200 | Human | Rabbit IgG | 1:200 | - | - |
| Desmin AF488 | Y66 | NA | Abcam | Ab185033 | Human | Rabbit IgG | 1:200 | <15 | - |
| E-cadherin AF647 | DECMA-1 | AB_2563955 | BioLegend | 147308 | Mouse | Rat IgG1, κ | 1:100 | <15 | - |
| EpCAM BV510 | G8.8 | AB_2738075 | BD Biosciences | 563216 | Mouse | Rat IgG2a, κ | 1:100 | - | <15 |
| EpCAM AF594 | G8.8 | AB_2563322 | BioLegend | 118222 | Mouse | Rat IgG2a, κ | 1:400 | >120 | - |
| EpCAM AF647 | G8.8 | AB_1134101 | BioLegend | 118212 | Mouse | Rat IgG2a, κ | 1:200 | <15 | - |
| EpCAM AF488 | 9C4 | AB_756084 | BioLegend | 324210 | Human | Mouse IgG2b, κ | 1:300 | <15 | - |
| EpCAM AF594 | 9C4 | AB_2563209 | BioLegend | 324228 | Human | Mouse IgG2b, κ | 1:500 | >120 | - |
| EpCAM AF647 | 9C4 | AB_1134101 | BioLegend | 324212 | Human | Mouse IgG2b, κ | 1:100 | <15 | - |
| EpCAM | - | AB_1603782 | Abcam | Ab71916 | Human | Rabbit IgG | 1:50 | - | - |
| F4/80 BV421 | BM8 | AB_11203717 | BioLegend | 123132 | Mouse | Rat IgG2a, κ | 1:50 | - | <15 |
| F4/80 PE | BM8 | AB_465923 | Thermo | 12-4801-82 | Mouse | Rat IgG2a, κ | 1:100 | <15 | - |
| F4/80 AF647 | BM8 | AB_893480 | BioLegend | 123122 | Mouse | Rat IgG2a, κ | 1:50 | <15 | - |
| F4/80 AF700 | BM8 | AB_2293450 | BioLegend | 123130 | Mouse | Rat IgG2a, κ | 1:50 | <15 | - |
| Fibronectin AF532 | 2F4 | NA | Novus | NBP2-22113AF532 | Human | Mouse IgG$_1$ | 1:25 | <15 | - |
| Fibrinogen | - | AB_10900171 | Abcam | Ab118533 (Unconjugated) | Human | Sheep IgG | 1:200 | <15 | - |
| Foxp3 AF488 | FJK-16s | AB_763537 | Thermo | 53-5773-82 | Mouse | Rat IgG2a, κ | 1:50 | <15 | - |
| Foxp3 AF532 | FJK-16s | AB_11218870 | Thermo | 58-5773-82 | Mouse | Rat IgG2a, κ | 1:50 | <15 | - |
| Foxp3 PE | FJK-16s | AB_465936 | Thermo | 12-5773-82 | Mouse | Rat IgG2a, κ | 1:50 | <15 | - |
| Foxp3 eF570 | FJK-16s | AB_11219073 | Thermo | 41-5773-82 | Mouse | Rat IgG2a, κ | 1:50 | <15 | - |
| Foxp3 eF660 | FJK-16s | AB_11218868 | Thermo | 50-5773-82 | Mouse | Rat IgG2a, κ | 1:50 | <15 | - |
| FOXP3 eF570 | 236A/E7 | AB_2573609 | Thermo | 41-4777-82 | Human | Mouse IgG$_1$, κ | 1:25 | <15 | - |
| GL-7 PE | GL7 | AB_10715834 | BD Biosciences | 561530 | Mouse | Rat IgM, κ | 1:100 | <15 | - |
| Glutamine synthetase | - | AB_880241 | Abcam | Ab49873 | Human, Mouse | Rabbit IgG | 1:200 | - | - |
| Glycophorin | EPR8200 | - | Abcam | Ab218372 | Human | Rabbit IgG | 1:300 | - | - |
| gp38 AF488 | 8.1.1 | AB_1133992 | BioLegend | 127405 | Mouse | Hamster IgG | 1:50 | <15 | - |
| HLA-DR AF488 | L243 | AB_493176 | BioLegend | 307619 | Human | Mouse IgG2a, κ | 1:200 | <15 | - |
| HLA-DR iF594 | L243 | NA | Caprico Biotechnologies | 1032136 | Human | Mouse IgG2a, κ | 1:100 | <15 | - |
| Hoechst 33342 | - | - | Biotium | 40046 | | - | 1:5,000 | - | - |
| ICOS AF488 | C398.4A | AB_2122584 | BioLegend | 313514 | Human | Hamster IgG | 1:25 | <15 | - |
| IgA AF555 | - | AB_2794378 | SouthernBiotech | 1040-32 | Human | Goat IgG | 1:500 | <15 | - |
| IgA1 AF647 | B3506B4 | AB_2796658 | SouthernBiotech | 9130-31 | Human | Mouse IgG$_1$, κ | 1:500 | <15 | - |
| IgA2 AF488 | A9604D2 | AB_2796665 | SouthernBiotech | 9140-30 | Human | Mouse IgG$_1$, κ | 1:500 | <15 | - |
| IgD AF488 | 11-26c.2a | AB_10730619 | BioLegend | 405718 | Mouse | Rat IgG2a, κ | 1:400 | <15 | - |
| IgD AF594 | 11-26c.2a | AB_2565572 | BioLegend | 405740 | Mouse | Rat IgG2a, κ | 1:400 | >120 | - |
| IgD AF700 | 11-26c.2a | AB_2563340 | BioLegend | 405729 | Mouse | Rat IgG2a, κ | 1:50 | <15 | - |
| IgD AF488 | IA6-2 | AB_11150397 | BioLegend | 348216 | Human | Mouse IgG2a, κ | 1:25 | <15 | - |
| IgD AF647 | IA6-2 | AB_2563269 | BioLegend | 348228 | Human | Mouse IgG2a, κ | 1:100 | <15 | - |
| IgM AF647 | EPR5539-65-4 | NA | Abcam | Ab200629 | Human | Rabbit mAb | 1:100 | <15 | - |
| IRF4 FITC | 3E4 | AB_2572538 | Thermo | 11-9858-82 | Human, Mouse | Rat IgG1, κ | 1:50 | <30 | - |
| IRF4 PE | 3E4 | AB_2563004 | Thermo | 12-9858-82 | Human, Mouse | Rat IgG1, κ | 1:50 | <15 | - |
| JOJO-1 | - | - | Thermo | J11372 | | - | 1:10,000 | - | - |
| Ki-67 AF488 | B56 | AB_647087 | BD Biosciences | 558616 | Human, Mouse | Mouse $IgG_1$, κ | 1:50 | <15 | - |
| Ki-67 AF700 | B56 | AB_10611571 | BD Biosciences | 561277 | Human, Mouse | Mouse $IgG_1$, κ | 1:50 | <15 | - |
| KLRG1 AF488 | 2F1 | AB_10898017 | BD Biosciences | 561619 | Mouse | Hamster $IgG_2$, κ | 1:50 | <15 | - |
| Laminin 1 + 2 | - | AB_305933 | Abcam | Ab7463 | Human, Mouse | Rabbit IgG | 1:100 | - | - |
| Langerin | 929F3.01 | NA | Novus | DDX0362P-100 | Human | Rat IgG2a | 1:50 | - | - |
| Lumican AF532 | - | AB_2139484 | R&D | AF2846 (Unconjugated) | Human | Goat IgG | 1:50 | <15 | - |
| Lumican AF700 | - | AB_2139484 | R&D | AF2846 (Unconjugated) | | Goat IgG | 1:25 | <15 | - |
| Ly-6G AF488 | 1A8 | AB_2561340 | BioLegend | 127626 | Mouse | Rat IgG2a, κ | 1:50 | <15 | - |
| Lysozyme | - | AB_303050 | Abcam | Ab2408 | Human | Rabbit IgG | 1:50 | - | - |
| Lyve-1 eF570 | ALY7 | AB_2573596 | Thermo | 41-0443-82 | Mouse | Rat IgG1, κ | 1:100 | <15 | - |
| Lyve-1 | - | AB_355144 | R&D | AF2089 (Unconjugated) | Human | Goat IgG | 1:100 | <15 | - |
| MARCO | - | AB_2643767 | Thermo | PA5-64134 | Human | Rabbit IgG | 1:25 | - | - |
| MHC-II BV421 | M5/114.15.2 | AB_10900075 | BioLegend | 107631 | Mouse | Rat IgG2b, κ | 1:400 | - | <15 |
| MHC-II AF647 | M5/114.15.2 | AB_493525 | BioLegend | 107618 | Mouse | Rat IgG2b, κ | 1:600 | <15 | - |
| MHC-II AF700 | M5/114.15.2 | AB_493727 | BioLegend | 107622 | Mouse | Rat IgG2b, κ | 1:100 | <15 | - |
| NF-H/NF-M AF488 | SM1-35 | AB_2750328 | BioLegend | 835614 | Human | Mouse $IgG_1$, κ | 1:50 | <15 | - |
| NK1.1 BV421 | PK136 | AB_10895916 | BioLegend | 108731 | Mouse | Mouse IgG2a, κ | 1:50 | - | <15 |
| p53 PE | PAb 240 | NA | Novus | NB200-103PE | Human | Mouse $IgG_1$, κ | 1:50 | <15 | - |
| Pax5 AF647 | 1H9 | AB_2562425 | BioLegend | 649704 | Mouse | Rat IgG2a, κ | 1:100 | <15 | - |
| PD-1 BV421 | 29F.1A.12 | AB_10900085 | BioLegend | 135217 | Mouse | Rat IgG2a, κ | 1:100 | - | <15 |
| PD-1 PE | 29F.1A.12 | AB_1877231 | BioLegend | 135206 | Mouse | Rat IgG2a, κ | 1:100 | <15 | - |
| PD-1 AF488 | EH12.2H7 | AB 2563594 | BioLegend | 329936 | Human | Mouse $IgG_1$, κ | 1:200 | <15 | - |
| PD-1 PE | EH12.2H7 | AB_940481 | BioLegend | 329906 | Human | Mouse $IgG_1$, κ | 1:200 | <15 | - |
| PD-1 AF647 | EH12.2H7 | AB_940471 | BioLegend | 329910 | Human | Mouse $IgG_1$, κ | 1:200 | <15 | - |
| Podocin | JB51-33 | NA | Novus | NBP2-75624 | Human | Rabbit IgG | 1:50 | - | - |
| RORγt | AFKJS-9 | AB_1834475 | Thermo | 14-6988-82 | Mouse | Rat IgG2a | 1:200 | - | - |
| SiglecF PE | E50-2440 | AB_394341 | BD Biosciences | 552126 | Mouse | Rat IgG2a, κ | 1:100 | <15 | - |
| SiglecF AF700 | 1RNM44N | AB_2637126 | Thermo | 56-1702-80 | Mouse | Rat IgG2a, κ | 1:50 | <15 | - |
| SIRPα AF488 | P84 | AB_2650815 | BioLegend | 144024 | Mouse | Rat IgG1, κ | 1:50 | <15 | - |
| SIRPα AF647 | P84 | AB_2721300 | BioLegend | 144027 | Mouse | Rat IgG1, κ | 1:200 | <15 | - |
| αSMA AF488 | 1A4 | AB_2574460 | Thermo | 53-9760-80 | Human, Mouse | Mouse IgG2a, κ | 1:500 | <15 | - |
| αSMA eF570 | 1A4 | AB_2573631 | Thermo | 41-9760-82 | Human, Mouse | Mouse IgG2a, κ | 1:300 | <15 | - |
| αSMA eF660 | 1A4 | AB_2574461 | Thermo | 53-9760-82 | Human, Mouse | Mouse IgG2a, κ | 1:500 | <15 | - |
| SPARC AF532 | - | AB_355728 | R&D | AF941 (Unconjugated) | Human | Goat IgG | 1:50 | <15 | - |
| Syndecan-1 | - | AB_442186 | R&D | AF2780 | Human | Goat IgG | 1:50 | - | - |
| TCRγδ AF488 | GL3 | AB_2562771 | BioLegend | 118128 | Mouse | Hamster IgG | 1:50 | <15 | - |
| TCRγδ PE | GL3 | AB_313832 | BioLegend | 118108 | Mouse | Hamster IgG | 1:100 | <15 | - |
| TCRγδ PE | B1 | AB_1089218 | BioLegend | 331210 | Human | Mouse $IgG_1$, κ | 1:100 | <15 | - |
| Tim-3 AF532 | 344823 | AB_2232900 | R&D | MAB2365 (Unconjugated) | Human | Rat IgG2a | 1:25 | <15 | - |
| Tim-3 AF647 | 344823 | AB_2232900 | R&D | MAB2365 (Unconjugated) | Human | Rat IgG2a | 1:50 | <15 | - |
| Tim-4 BV421 | 21H12 | AB_2741037 | BD Biosciences | 742773 | Mouse | Rat IgG1, κ | 1:100 | - | <15 |
| TMPRSS2 | EPR3861 | AB_10585592 | Abcam | Ab92323 | Human | Rabbit IgG | 1:50 | - | - |
| Tryptase | AA1 | AB_303023 | Abcam | Ab2378 | Human | Mouse IgG$_1$ | 1:50 | - | - |
| Uromodulin FITC | 10.32 | AB_10012622 | Novus | NBP1-50431 | Human | Mouse IgG2b | 1:25 | <30 | - |
| TCR V$\alpha$7.2 AF647 | 3C10 | AB_2566335 | BioLegend | 351726 | Human | Mouse IgG$_1$, κ | 1:50 | <15 | - |
| Vimentin AF532 | O91D3 | NA | BioLegend | Custom | Human | Mouse IgG2a | 1:200 | <15 | - |
| Vimentin AF594 | O91D3 | AB_2566179 | BioLegend | 677804 | Human | Mouse IgG2a | 1:300 | >120 | - |
| Vimentin AF647 | O91D3 | AB_2616801 | BioLegend | 677807 | Human | Mouse IgG2a | 1:600 | <15 | - |
| anti-chicken IgY FITC | - | AB_923386 | Thermo | SA1-72000 | - | Donkey IgG | 1:200 | <30 | - |
| anti-chicken IgY AF555 | - | AB_2762844 | Thermo | A32932 | - | Goat IgG | 1:400 | <15 | - |
| anti-goat IgG AF488 | - | AB_2534102 | Thermo | A-11055 | - | Donkey IgG | 1:400 | <15 | - |
| anti-goat IgG AF555 | - | AB_2535853 | Thermo | A21432 | - | Donkey IgG | 1:300 | <15 | - |
| anti-goat IgG AF680 | - | AB_2535741 | Thermo | A21084 | - | Donkey IgG | 1:300 | <15 | - |
| anti-hamster IgG AF647 | - | AB_2535868 | Thermo | A-21451 | - | Goat IgG | 1:400 | <15 | - |
| anti-mouse IgM AF488 | - | AB_2340844 | Jackson Immunoresearch | 715-545-020 | - | Donkey IgG | 1:200 | <15 | - |
| anti-rabbit IgG AF488 | - | AB_2576217 | Thermo | A-11034 | - | Goat IgG | 1:400 | <15 | - |
| anti-rabbit IgG AF532 | - | AB_2534076 | Thermo | A-11009 | - | Goat IgG | 1:400 | <15 | - |
| anti-rabbit IgG AF555 | - | AB_141784 | Thermo | A-21428 | - | Goat IgG | 1:400 | <15 | - |
| anti-rabbit IgG AF594 | - | AB_2534095 | Thermo | A-11037 | - | Goat IgG | 1:400 | >120 | - |
| anti-rabbit IgG AF647 | - | AB_2535813 | Thermo | A-21245 | - | Goat IgG | 1:400 | <15 | - |
| anti-rabbit IgG AF700 | - | AB_2535709 | Thermo | A-21038 | - | Goat IgG | 1:400 | <15 | - |
| anti-rabbit IgG AF750 | - | AB_2535710 | Thermo | A-21039 | - | Goat IgG | 1:400 | <15 | - |
| anti-rabbit IgG AF488 | - | AB_2535792 | Thermo | A-21206 | - | Donkey IgG | 1:300 | <15 | - |
| anti-rabbit IgG AF555 | - | AB_162543 | Thermo | A-31572 | - | Donkey IgG | 1:300 | <15 | - |
| anti-rabbit IgG AF647 | - | AB_2536183 | Thermo | A-31573 | - | Donkey IgG | 1:400 | <15 | - |
| anti-rabbit IgG AF488 | - | AB_2572214 | Thermo | Z-25302 | - | Goat Fab$_2$ | - | <15 | - |
| anti-rabbit IgG AF532 | - | AB_2736945 | Thermo | Z-25303 | - | Goat Fab$_2$ | - | <15 | - |
| anti-rabbit IgG AF555 | - | AB_2736950 | Thermo | Z-25305 | - | Goat Fab$_2$ | - | <15 | - |
| anti-rabbit IgG AF594 | - | AB_2736956 | Thermo | Z-25307 | - | Goat Fab$_2$ | - | >120 | - |
| anti-rabbit IgG AF647 | - | AB_2736962 | Thermo | Z-25308 | - | Goat Fab$_2$ | - | <15 | - |
| anti-rat IgG AF647 | - | AB_141778 | Thermo | A-21247 | - | Goat IgG | 1:400 | <15 | - |
| anti-rat IgG AF647 | - | AB_2340694 | Jackson Immunoresearch | 712-605-153 | - | Donkey IgG | 1:200 | <15 | - |
| anti-sheep IgG AF680 | - | AB_2535755 | Thermo | A-21102 | - | Donkey IgG | 1:300 | <15 | - |

## Table S2. IBEX panels for human organs.

## Lymph node (Manual IBEX): 38 parameters, 9 cycles

| Cycle | Marker | Clone | Conjugate | Isotype | Vendor | Catalog Number | Dilution |
|---|---|---|---|---|---|---|---|
| 1 | Hoechst | - | - | | Biotium | 40046 | 1:5000 |
| | CD20 | L26 | AF488 | Mouse IgG2b, κ | eBioscience | 53-0202-82 | 1:100 |
| | CD10 | FR4D11 | PE | Mouse IgG1, κ | Caprico Biotechnologies | 103924 | 1:25 |
| | CD3 | UCHT1 | AF594 | Mouse IgG1, κ | BioLegend | 300446 | 1:50 |
| | BCL2 | 100 | AF647 | Mouse IgG1 | BioLegend | 658705 | 1:25 |
| | Collagen IV | Polyclonal rabbit IgG | Goat anti-Rabbit IgG AF700 | Rabbit IgG<br>Goat IgG | Abcam<br>Thermo | Ab6586<br>A-21038 | 1:50<br>1:250 |
| 2 | IgD | IA6-2 | AF488 | Mouse IgG2a, κ | BioLegend | 348215 | 1:25 |
| | CD23 | EBVCS-5 | AF532 | Mouse IgG1, κ | BioLegend | Custom | 1:25 |
| | CD138 | MI15 | PE | Mouse IgG1, κ | BioLegend | 356503 | 1:200 |
| | CD3 | UCHT1 | AF594 | Mouse IgG1, κ | BioLegend | 300446 | 1:100 |
| | BCL6 | K112-91 | AF647 | Mouse IgG1, κ | BD Biosciences | 561525 | 1:10 |
| | CD31 | WM59 | AF700 | Mouse IgG1, κ | BioLegend | 303133 | 1:25 |
| 3 | HLA-DR | L243 | AF488 | Mouse IgG2a, κ | BioLegend | 307619 | 1:100 |
| | CD21 | Bu32 | AF532 | Mouse IgG1, κ | BioLegend | NA, Custom | 1:600 |
| | CD1c | L161 | PE | Mouse IgG1, κ | BioLegend | 331505 | 1:50 |
| | CD3 | UCHT1 | AF594 | Mouse IgG1, κ | BioLegend | 300446 | 1:100 |
| | CD163 | GH1/61 | AF647 | Mouse IgG1, κ | BioLegend | 333620 | 1:100 |
| | CD11c | B-Ly6 | AF700 | Mouse IgG1, κ | BD Biosciences | 561352 | 1:25 |
| 4 | CD8 | SK1 | AF488 | Mouse IgG1, κ | BioLegend | 344716 | 1:50 |
| | FOXP3 | 236A/E7 | eF570 | Mouse IgG1, κ | eBioscience | 41-4777-82 | 1:50 |
| | CD3 | UCHT1 | AF594 | Mouse IgG1, κ | BioLegend | 300446 | 1:50 |
| | CD25 | M-A251 | AF647 | Mouse IgG1, κ | BioLegend | 356127 | 1:50 |
| 5 | CD4 | RPA-T4 | FITC | Mouse IgG1, κ | Thermo | 58-0049-41 | 1:25 |
| | SPARC | Goat IgG | AF532 | Goat IgG | R&D | AF941 | 1:25 |
| | PD-1 | EH12.2H7 | PE | Mouse IgG1, κ | BioLegend | 329905 | 1:200 |
| | CD3 | UCHT1 | AF594 | Mouse IgG1, κ | BioLegend | 300446 | 1:50 |
| | CD69 | FN50 | AF647 | Mouse IgG1, κ | BioLegend | 310918 | 1:25 |
| | CD38 | HIT2 | AF700 | Mouse IgG1, κ | BioLegend | 303524 | 1:25 |
| 6 | ICOS | C398.4A | AF488 | Hamster IgG | BioLegend | 313514 | 1:25 |
| | CXCL13 | Goat IgG | AF532 | Goat IgG | R&D | AF801 (Unconjugated) | 1:25 |
| | IRF4 | IRF4.3E4 | PE | Rat IgG1, κ | BioLegend | 646403 | 1:50 |
| | CD3 | UCHT1 | AF594 | Mouse IgG1, κ | BioLegend | 300446 | 1:50 |
| | CD68 | KP1 | AF647 | Mouse IgG1, κ | Santa Cruz | sc-20060 | 1:100 |
| | Ki-67 | B56 | AF700 | Mouse IgG1, κ | BD Biosciences | 561277 | 1:50 |
| 7 | CD44 | IM7 | AF532 | Rat IgG2b, κ | BioLegend | Custom | 1:50 |
| | CD35 | E11 | PE | Mouse IgG1, κ | BioLegend | 333406 | 1:800 |
| | CD3 | UCHT1 | AF594 | Mouse IgG1, κ | BioLegend | 300446 | 1:100 |
| | Vα7.2 | 3C10 | AF647 | Mouse IgG1, κ | BioLegend | 351726 | 1:50 |
| 8 | SMA-alpha | 1A4 | AF488 | Mouse IgG2a, κ | Thermo | 53-9760-80 | 1:100 |
| | CD106 | STA | PE | Mouse IgG1, κ | BioLegend | 305805 | 1:50 |
| | CD3 | UCHT1 | AF594 | Mouse IgG1, κ | BioLegend | 300446 | 1:50 |
| 9 | Clec9a | 8F9 | AF488 | Mouse IgG2a, κ | BioLegend | Custom | 1:25 |
| | Lyve-1 | Polyclonal goat IgG | AF532 | Goat IgG | R&D | AF2089 (Unconjugated) | 1:100 |
| | CD45 | F10-89-4 | PE-iF594 | Mouse IgG2a, κ | Caprico Biotechnologies | 1016185 | 1:100 |
| | CD3 | UCHT1 | AF594 | Mouse IgG1, κ | BioLegend | 300446 | 1:50 |
| | DC-SIGN | 9E9A8 | AF647 | Mouse IgG2a, κ | BioLegend | 330112 | 1:50 |

## Spleen (Manual IBEX): 25 parameters, 4 cycles

| Cycle | Marker | Clone | Conjugate | Isotype | Vendor | Catalog Number | Dilution |
|---|---|---|---|---|---|---|---|
| 1 | Hoechst | - | - | | Biotium | 40046 | 1:5000 |
| | CD20 | L26 | AF488 | Mouse IgG2b, κ | eBioscience | 53-0202-82 | 1:200 |
| | CD21 | Bu32 | AF532 | Mouse IgG1, κ | BioLegend | NA, Custom | 1:400 |
| | Glycophorin AF555 | EPR8200 | AF555 | Rabbit IgG | Abcam<br>Abcam | Ab218372<br>Ab269820 | 1:300 |
| | CD68 | KP1 | iF594 | Mouse IgG1, κ | Caprico Biotechnologies | 1064135 | 1:50 |
| | CD54 | HA58 | AF647 | Mouse IgG1, κ | BioLegend | 353114 | 1:100 |
| | Fibrinogen | Polyclonal sheep IgG<br>Polyclonal donkey IgG | Donkey anti-sheep IgG AF680 | Sheep IgG<br>Donkey IgG | Abcam<br>Thermo | Ab118533<br>A-21102 | 1:200<br>1:400 |
| 2 | CD15 | MMA<br>Donkey anti-mouse IgM | Donkey anti-mouse IgM AF488 | Mouse IgM, κ<br>Donkey IgG | BD Biosciences<br>Jackson Immunoresearch | 347420<br>715-545-020 | 1:50<br>1:200 |
| | CD163 | GH1/61 | AF532 | Mouse IgG1, κ | BioLegend | Custom | 1:200 |
| | CD138 | MI15 | PE | Mouse IgG1, κ | BioLegend | 356503 | 1:200 |
| | HLA-DR | L243 | iF594 | Mouse IgG2a, κ | Caprico Biotechnologies | 1032136 | 1:100 |
| | CD49a | TS2/7 | AF647 | Mouse IgG1, κ | BioLegend | 328310 | 1:200 |
| | CD11c | B-Ly6 | AF700 | Mouse IgG1, κ | BD Biosciences | 561352 | 1:25 |
| 3 | CD8 | SK1 | AF488 | Mouse IgG1, κ | BioLegend | 344716 | 1:25 |
| | Vimentin | O91D3 | AF532 | Mouse IgG2a, κ | BioLegend | Custom | 1:200 |
| | CD31 | WM59 | PE | Mouse IgG1, κ | BioLegend | 303106 | 1:100 |
| | CD3 | UCHT1 | iF594 | Mouse IgG1, κ | AAT Bioquest | 103400C0 | 1:50 |
| | CD4 | RPA-T4 | AF647 | Mouse IgG1, κ | BioLegend | 300520 | 1:100 |
| | Ki-67 | B56 | AF700 | Mouse IgG1, κ | BD Biosciences | 561277 | 1:25 |
| 4 | CD61 | Y2/51 | FITC | Mouse IgG1, κ | Miltenyi | 130-098-682 | 1:25 |
| | SPARC | Goat IgG | AF532 | Goat IgG | R&D | AF941 | 1:50 |
| | SMA-alpha | 1A4 | eF570 | Mouse IgG2a, κ | Thermo | 41-9760-82 | 1:300 |
| | CD45 | F10-89-4 | iF594 | Mouse IgG2a, κ | Caprico Biotechnologies | 1016138 | 1:100 |
| | CD44 | IM7 | AF647 | Rat IgG2b, κ | BioLegend | 103018 | 1:200 |
| | Lumican | Polyclonal goat IgG | AF680 | Goat IgG | R&D<br>Thermo | AF2846<br>A-20188 | 1:25 |

## Human Liver (Manual IBEX): 22 parameters, 4 cycles

| Cycle | Marker | Clone | Conjugate | Isotype | Vendor | Catalog Number | Dilution |
|---|---|---|---|---|---|---|---|
| 1 | Hoechst | - | - | | Biotium | 40046 | 1:5000 |
| | Keratin 7 | W16155A | AF488 | Rat IgG2a, κ | BioLegend | 601605 | 1:50 |
| | ASS1 | Polyclonal rabbit IgG<br>Polyclonal donkey IgG | Donkey anti-rabbit IgG AF555 | Rabbit IgG<br>Donkey IgG | Abcam<br>Thermo | Ab170952<br>A31572 | 1:100<br>1:300 |
| | CD34 | 4H11 | iF594 | Mouse IgG1 | AAT Bioquest | 103400C0 | 1:50 |
| | CD54 | HA58 | AF647 | Mouse IgG1, κ | BioLegend | 353114 | 1:50 |
| | Lyve-1 | Polyclonal goat IgG<br>Polyclonal donkey IgG | Donkey anti-goat IgG AF555 | Goat IgG<br>Donkey IgG | R&D<br>Thermo | AF2086-SP<br>A21084 | 1:400<br>1:300 |
| 2 | HLA-DR | L243 | AF488 | Mouse IgG2a, κ | BioLegend | 307619 | 1:100 |
| | CD163 | GH1/61 | AF532 | Mouse IgG1, κ | BioLegend | 99486 | 1:100 |
| | CD138 | MI15 | PE | Mouse IgG1, κ | BioLegend | 356503 | 1:200 |
| | CD68 | KP1 | iF594 | Mouse IgG1, κ | Caprico Biotechnologies | 1064136 | 1:100 |
| | Fibrinogen | Polyclonal sheep IgG<br>Polyclonal donkey IgG | Donkey anti-sheep IgG AF680 | Sheep IgG<br>Donkey IgG | Abcam<br>Thermo | Ab118533<br>A-21102 | 1:100<br>1:300 |
| 3 | CD8 | SK1 | AF488 | Mouse IgG1, κ | BioLegend | 344716 | 1:25 |
| | B-Tubulin 3 | TUJ1 | AF532 | Mouse IgG2a, κ | BioLegend | Custom | 1:100 |
| | Glutamine Synthetase | Polyclonal rabbit IgG<br>Anti-rabbit Fab | AF555 | Rabbit IgG | Abcam<br>Thermo | Ab49873<br>Z-25303 | 1:100 |
| | CD3 | UCHT1 | iF594 | Mouse IgG1, κ | Caprico Biotechnologies | 1053134 | 1:50 |
| | CD4 | RPA-T4 | AF647 | Mouse IgG1, κ | BioLegend | 300520 | 1:50 |
| | Ki-67 | B56 | AF700 | Mouse IgG1, κ | BD Biosciences | 561277 | 1:25 |
| 4 | αSMA | 1A4 | AF488 | Mouse IgG2a, κ | Thermo | 53-9760-80 | 1:600 |
| | CD44 | IM7 | AF532 | Rat IgG2b, κ | BioLegend | Custom | 1:50 |
| | CD49a | TS2/7 | PE | Mouse IgG1, κ | BioLegend | 328303 | 1:500 |
| | CD45 | HI30 | AF594 | Mouse IgG2a, κ | BioLegend | 304060 | 1:100 |
| | Vimentin | O91D3 | AF647 | Mouse IgG2a, κ | BioLegend | 677807 | 1:600 |

## Lymph node (Automated IBEX): 24 markers, 6 cycles

| Cycle | Marker | Clone | Conjugate | Isotype | Vendor | Catalog Number | Dilution |
|---|---|---|---|---|---|---|---|
| 1 | Hoechst | - | - | | Biotium | 40046 | 1:5000 |
| | CD20 | L26 | AF488 | Mouse IgG2b, κ | eBioscience | 53-0202-82 | 1:100 |
| | CD34 | 4H11 | iF594 | Mouse IgG1 | AAT Bioquest | 103400C0 | 1:25 |
| | CD54 | HA58 | AF647 | Mouse IgG1, κ | BioLegend | 353114 | 1:50 |
| | Lyve-1 | Polyclonal goat IgG | | Goat IgG | R&D | AF2086-SP | 1:400 |
| | Collagen IV | Polyclonal rabbit IgG | | Rabbit IgG | Abcam | Ab6586 | 1:50 |
| 2 | Collagen IV | Donkey anti-rabbit IgG | AF488 | Donkey IgG | Thermo | A-21206 | 1:200 |
| | Lyve-1 | Donkey anti-goat IgG | AF555 | Donkey IgG | Thermo | A-21432 | 1:200 |
| | CD68 | KP1 | iF594 | Mouse IgG1, κ | Caprico Biotechnologies | 1064135 | 1:50 |
| | CD49a | TS2/7 | AF647 | Mouse IgG1, κ | BioLegend | 328310 | 1:50 |
| 3 | CD21 | Bu32 | AF488 | Mouse IgG1, κ | BioLegend | 98148 | 1:100 |
| | CD35 | E11 | PE | Mouse IgG1, κ | BioLegend | 333406 | 1:200 |
| | HLA-DR | L243 | iF594 | Mouse IgG2a, κ | Caprico Biotechnologies | 1032136 | 1:50 |
| | CD163 | GH1I/61 | AF647 | Mouse IgG1, κ | BioLegend | 333620 | 1:50 |
| 4 | CD8 | SK1 | AF488 | Mouse IgG1, κ | BioLegend | 344716 | 1:25 |
| | PD-1 | EH12.2H7 | PE | Mouse IgG1, κ | BioLegend | 329906 | 1:50 |
| | CD3 | UCHT1 | iF594 | Mouse IgG1, κ | Caprico Biotechnologies | 1053135 | 1:50 |
| | CD4 | RPA-T4 | AF647 | Mouse IgG1, κ | BioLegend | 300520 | 1:25 |
| 5 | Desmin | Y66 | AF488 | Rabbit IgG | Abcam | Ab185033 | 1:50 |
| | CD31 | WM59 | PE | Mouse IgG1, κ | BioLegend | 303106 | 1:100 |
| | CD45 | F10-89-4 | iF594 | Mouse IgG2a, κ | Caprico Biotechnologies | 1016136 | 1:100 |
| | CD138 | MI15 | AF647 | Mouse IgG1, κ | BioLegend | 356524 | 1:100 |
| 6 | CD44 | IM7 | AF488 | Rat IgG2b, κ | BioLegend | 103016 | 1:50 |
| | CD94 | DX22 | PE | Mouse IgG1, κ | BioLegend | 305506 | 1:50 |
| | Vimentin | O91D3 | AF594 | Mouse IgG2a, κ | BioLegend | 677804 | 1:300 |
| | αSMA | 1A4 | eF660 | Mouse IgG2a, κ | Thermo | 53-9760-82 | 1:200 |

## Jejunum (Automated IBEX): 24 markers, 6 cycles

| Cycle | Marker | Clone | Conjugate | Isotype | Vendor | Catalog Number | Dilution |
|---|---|---|---|---|---|---|---|
| 1 | Hoechst | - | - | | Biotium | 40046 | 1:5000 |
| | EpCAM | 9C4 | AF488 | Mouse IgG2b, κ | BioLegend | 324210 | 1:300 |
| | CD106 | STA | PE | Mouse IgG1, κ | BioLegend | 305806 | 1:50 |
| | CD34 | 4H11 | iF594 | Mouse IgG1 | AAT Bioquest | 103400C0 | 1:25 |
| | CD54 | HA58 | AF647 | Mouse IgG1, κ | BioLegend | 353114 | 1:50 |
| | Lysozyme | Polyclonal rabbit IgG | | Rabbit IgG | Abcam | Ab2408 | 1:50 |
| | Lyve-1 | Polyclonal goat IgG | | Goat IgG | R&D | AF2089 | 1:100 |
| 2 | Lysozyme | Donkey anti-rabbit IgG | AF488 | Donkey IgG | Thermo | A-21206 | 1:200 |
| | Lyve-1 | Donkey anti-goat IgG | AF555 | Donkey IgG | Thermo | A-21432 | 1:200 |
| | CD68 | KP1 | iF594 | Mouse IgG1, κ | Caprico Biotechnologies | 1064135 | 1:50 |
| | CD49a | TS2/7 | AF647 | Mouse IgG1, κ | BioLegend | 328310 | 1:50 |
| 3 | Tubulin3 | TUJ1 | AF488 | Mouse IgG2a, κ | BioLegend | 801203 | 1:50 |
| | CD31 | WM59 | PE | Mouse IgG1, κ | BioLegend | 303106 | 1:100 |
| | HLA-DR | L243 | iF594 | Mouse IgG2a, κ | Caprico Biotechnologies | 1032136 | 1:50 |
| | Cytokeratin | AE1/AE3 | eF660 | Mouse IgG1 | Thermo | 50-9003-82 | 1:50 |
| 4 | CD8 | SK1 | AF488 | Mouse IgG1, κ | BioLegend | 344716 | 1:25 |
| | PD-1 | EH12.2H7 | PE | Mouse IgG1, κ | BioLegend | 329906 | 1:50 |
| | CD3 | UCHT1 | iF594 | Mouse IgG1, κ | Caprico Biotechnologies | 1053135 | 1:50 |
| | CD4 | RPA-T4 | AF647 | Mouse IgG1, κ | BioLegend | 300520 | 1:25 |
| 5 | IgA2 | A9604D2 | AF488 | Mouse IgG1, κ | SouthernBiotech | 9140-30 | 1:400 |
| | CD138 | MI15 | PE | Mouse IgG1, κ | BioLegend | 356503 | 1:100 |
| | CD45 | F10-89-4 | iF594 | Mouse IgG2a, κ | Caprico Biotechnologies | 1016136 | 1:100 |
| | IgA1 | B3506B4 | AF647 | Mouse IgG1, κ | SouthernBiotech | 9130-31 | 1:400 |
| 6 | CD44 | IM7 | AF488 | Rat IgG2b, κ | BioLegend | 103016 | 1:100 |
| | αSMA | 1A4 | eF570 | Mouse IgG2a, κ | Thermo | 41-9760-82 | 1:200 |
| | Vimentin | O91D3 | AF647 | Mouse IgG2a, κ | BioLegend | 677807 | 1:300 |

## Skin (Automated IBEX): 19 markers, 5 cycles

| Cycle | Marker | Clone | Conjugate | Isotype | Vendor | Catalog Number | Dilution |
|---|---|---|---|---|---|---|---|
| 1 | Hoechst | - | - | | Biotium | 40046 | 1:5000 |
| | CD34 | 4H11 | iF594 | Mouse IgG1 | AAT Bioquest | 103400C0 | 1:25 |
| | CD54 | HA58 | AF647 | Mouse IgG1, κ | BioLegend | 353114 | 1:50 |
| | Lumican | Polyclonal goat IgG | | Goat IgG | R&D | AF2846 | 1:200 |
| | EpCAM | Polyclonal rabbit IgG | | Rabbit IgG | Abcam | Ab71916 | 1:50 |
| 2 | Lumican | Donkey anti-goat IgG | AF488 | Donkey IgG | Thermo | A-11055 | 1:400 |
| | EpCAM | Donkey anti-rabbit IgG | AF555 | Donkey IgG | Thermo | A-31572 | 1:200 |
| | CD68 | KP1 | iF594 | Mouse IgG1, κ | Caprico Biotechnologies | 1064135 | 1:50 |
| | CD49a | TS2/7 | AF647 | Mouse IgG1, κ | BioLegend | 328310 | 1:50 |
| | Langerin | 929F3.01 | | Rat IgG2a | Novus | DDX0362P-100 | 1:50 |
| 3 | HLA-DR | L243 | AF488 | Mouse IgG2a, κ | BioLegend | 307620 | 1:50 |
| | CD117 | 104D2 | PE | Mouse IgG1, κ | BioLegend | 313204 | 1:50 |
| | CD3 | UCHT1 | iF594 | Mouse IgG1, κ | Caprico Biotechnologies | 1053135 | 1:25 |
| | Langerin | Donkey anti-Rat IgG | AF647 | Donkey IgG | Jackson Immuno Research | 712-605-153 | 1:200 |
| 4 | CD44 | IM7 | AF488 | Rat IgG2b, κ | BioLegend | 103054 | 1:50 |
| | CD31 | WM59 | PE | Mouse IgG1, κ | BioLegend | 303106 | 1:100 |
| | CD45 | F10-89-4 | iF594 | Mouse IgG2a, κ | Caprico Biotechnologies | 1016136 | 1:100 |
| | CD138 | MI15 | AF647 | Mouse IgG1, κ | BioLegend | 356524 | 1:100 |
| | Keratin 14 | Poly9060 | | Chicken IgY | BioLegend | 906004 | 1:50 |
| 5 | Keratin 10 | DE-K10 | AF488 | Mouse IgG1, κ | Novus | NBP2-54402AF488 | 1:200 |
| | Keratin 14 | Goat anti-Chicken IgY | AF555 | Goat IgG | Invitrogen | A32932 | 1:400 |
| | Vimentin | O91D3 | AF594 | Mouse IgG2a, κ | BioLegend | 677807 | 1:300 |
| | αSMA | 1A4 | eF660 | Mouse IgG2a, κ | Thermo | 41-9760-82 | 1:50 |

## Kidney (Automated IBEX, FFPE): 16 markers, 5 cycles

| Cycle | Marker | Clone | Conjugate | Isotype | Vendor | Catalog Number | Dilution |
|---|---|---|---|---|---|---|---|
| 1 | Hoechst | - | - | | Biotium | 40046 | 1:5000 |
| | Cathepsin L | 204101 | | Rat IgG2a | R&D | MAB-9521 | 1:50 |
| | CD10 | Polyclonal goat IgG | | Goat IgG | R&D | AF1182 | 1:200 |
| | Podocin | JB51-33 | | Rabbit IgG | Novus | NBP2-75624 | 1:50 |
| | Cytokeratin | AE1/AE3 | AF488 | Mouse IgG1 | Thermo | 53-9003-82 | 1:50 |
| | CD34 | QBEND-10 | PE | Mouse IgG1, κ | Thermo | MAI-10205 | 1:50 |
| | CD68 | KP1 | iF594 | Mouse IgG1, κ | Caprico Biotechnologies | 1064135 | 1:25 |
| | αSMA | 1A4 | eF660 | Mouse IgG2a, κ | Thermo | 50-9760-82 | 1:50 |
| 2 | CD10 | Donkey anti-goat IgG | AF488 | Donkey IgG | Thermo | A-11055 | 1:400 |
| | Podocin | Donkey anti-rabbit IgG | AF555 | Donkey IgG | Thermo | A-21432 | 1:400 |
| | CD66b | G10F5 | iF594 | Mouse IgM, κ | Caprico Biotechnologies | 1075139 | 1:25 |
| | Cathepsin L | Donkey anti-rat IgG | AF647 | Rat IgG2a | Jackson Immunoresearch | 712-605-153 | 1:100 |
| | CD15 | MMA | | Mouse IgM, κ | BD | 347420 | 1:25 |
| 3 | CD15 Donkey anti-mouse IgM AF488 | Polyclonal | AF488 | Donkey IgG | Jackson Immunoresearch | 715-545-020 | 1:200 |
| | ACE2 | Polyclonal, Custom conjugate | AF555 | Goat IgG | R&D Abcam | AF933 Ab269820 | 1:50 |
| | SPARC | Polyclonal, Custom conjugate | iF594 | Goat IgG | R&D AAT Bioquest | AF941 1230 | 1:50 |
| | TMPRSS2 | EPR3861, Custom conjugate | AF647 | Rabbit IgG | Abcam Abcam | Ab92323 Ab269823 | 1:50 |
| 4 | Lumican iF594 | Polyclonal goat IgG<br>Custom conjugate | iF594 | Goat IgG | R&D<br>AAT Bioquest | AF2846<br>1230 | 1:200 |
| 5 | Uromodulin | 10.32 | FITC | Mouse IgG2b | Novus | NBP1-50431 | 1:25 |
| | Glycophorin | EPR8200,<br>Custom conjugate | AF555 | Rabbit IgG | Abcam<br>Abcam | Ab218372<br>Ab269820 | 1:300 |
| | Syndecan-1 | Polyclonal,<br>Custom conjugate | iF594 | Goat IgG | R&D<br>AAT Bioquest | AF2780<br>1230 | 1:50 |
| | Vimentin | O91D3 | AF647 | Mouse IgG2a, κ | BioLegend | 677807 | 1:200 |

Please see original IBEX publication for additional panels including the following mouse tissues: lymph node, spleen, thymus, lung, small intestine, and liver: https://doi.org/10.1073/pnas.2018488117.

## Table S3. Recommendations for optimal antibody and fluorophore pairing for IBEX panel design.

| Ex/Em Max (nm) | Fluorophore | Inactivation Method | Dim (1)-Brightest (5) | IBEX Method | Special Considerations |
|---|---|---|---|---|---|
| 350/461 | Hoechst 33342 | - | 5 | Both | Preferred fiducial for registration and nuclear segmentation<br>Overcomes high autofluorescence in this channel, especially in human tissues |
| 410/455 | Pacific Blue | <15 minutes | 1 | Manual (tested) | Dim, limited availability of direct conjugates<br>Use Hoechst 33342 in this channel |
| 405/421 | Brilliant Violet 421 | <15 minutes + Light | 4 | Manual (tested) | Use Hoechst 33342 in this channel<br>High autofluorescence in human tissues<br>Light inactivation is tedious for large areas |
| 405/510 | Brilliant Violet 510 | <15 minutes + Light | 1 | Manual (tested) | Use Hoechst 33342 in this channel<br>High autofluorescence in human tissues<br>Light inactivation is tedious for large areas |
| 490/525 | AF488 | <15 minutes | 4 | Both | Excellent fluorophore for IBEX<br>Wide availability of direct conjugates, labeling kits, and secondary antibodies<br>Pair with moderate to highly expressed markers |
| 490/525 | FITC | <30 minutes | 3 | Both | Prioritize AF488 over FITC conjugated antibodies |
| 532/554 | AF532 | <15 minutes | 3 | Manual | Suitable fluorophore for manual IBEX<br>Limited availability of direct conjugates and secondary antibodies<br>Pair with highly expressed markers and be mindful of spectral overlap between AF555/PE/eF570 channel |
| 533/544 | JOJO-1 | - | 4 | Manual | Avoid and use Hoechst 33342 as nuclear marker and fiducial instead |
| 555/580 | AF555 | <15 minutes | 3 | Both | Excellent fluorophore for IBEX<br>Wide availability of direct conjugates, labeling kits, and secondary antibodies<br>Pair with moderate to highly expressed markers and be mindful of spectral overlap with AF532 channel |
| 546/585 | eF570 | <15 minutes | 3 | Both | Good fluorophore for IBEX<br>Limited availability of direct conjugates, labeling kits, and secondary antibodies<br>Pair with moderate to highly expressed markers and be mindful of spectral overlap with AF532 channel |
| 488-561/565 | PE | <15 minutes | 5 | Both | Prioritize AF555 and eF570 conjugates<br>Prone to photobleaching but can be paired with dim markers<br>Be mindful of spectral overlap with AF532 channel and non-specific binding to tissues |
| 587/603 | iF594 | <15 minutes | 4 | Both | Excellent fluorophore for IBEX<br>Limited availability of direct conjugates, labeling kits, and secondary antibodies |
| 590/617 | AF594 | - | 5 | Both | Can be paired with a structural marker and used as fiducial or placed in the final IBEX cycle<br>Avoid and use Hoechst 33342 as fiducial and iF594 instead |
| 560/645 | eF615 | - | 4 | Both | Can be paired with a structural marker and used as fiducial or placed in the final IBEX cycle<br>Avoid and use Hoechst 33342 as fiducial and iF594 instead<br>Limited availability of direct conjugates, labeling kits, and secondary antibodies |
| 650/665 | AF647 | <15 minutes | 5 | Both | Excellent fluorophore for IBEX<br>Wide availability of direct conjugates, labeling kits, and secondary antibodies<br>Pair with dim to moderately expressed markers |
| 633/669 | eF660 | <15 minutes | 4 | Both | Excellent fluorophore for IBEX<br>Limited availability of direct conjugates, labeling kits, and secondary antibodies<br>Pair with dim to moderately expressed markers |
| 679/702 | AF680 | <15 minutes | 3 | Manual | Suitable fluorophore for manual IBEX, cannot be separated from adjacent channels on THUNDER<br>Limited availability of direct conjugates and secondary antibodies<br>Pair with highly expressed markers |
| 702/723 | AF700 | <15 minutes | 1 | Manual | Suitable fluorophore for manual IBEX<br>Limited availability of direct conjugates and secondary antibodies<br>Pair with highly expressed markers and place in Cycles 1-4 |
| 749/775 | AF750 | <15 minutes | 3 | Automated | Suitable fluorophore for automated IBEX on THUNDER but signal greatly diminished with AFC, in general avoid<br>Limited availability of direct conjugates and secondary antibodies |

## Table S4. Estimated costs for IBEX implementation.

| | |
|---|---|
| **Tissue Grossing and Processing**<br>• Reagents and consumables for 100 samples <$5,000<br>• Tools for embedding/processing for 100 samples <$1,000<br>. Cryostat (Leica CM1950) or equivalent, available in most core facilities ~$45,000<br>. Rotary microtome (Leica, RM2255) or equivalent, available in most core facilities <$20,000<br>• Stereomicroscope and illuminator <$5,000 | **Manual and Automated IBEX** |
| **Immunolabeling and microscopy**<br>• Leica TCS SP8 X confocal microscope or equivalent, highly variable $500,000-700,000<br>• PELCO BioWave Pro microwave and SteadyTemp Pro, optional (incubate at 37°C) ~$31,000<br>• ~$4 per antibody per run (2 µg in 400 µl), scales with complexity of panels and number of samples | **Manual IBEX** |
| **Immunolabeling and microscopy**<br>• Leica THUNDER microscope or equivalent, dependent on configuration $175,000-200,000<br>• ARIA Fluidics Device <$15,000<br>• Imaging chamber, vacuum, and heater from Warner instruments <$5,000<br>• Additional accessories (triggering cables) <$100<br>• ~$4 per antibody per run (2 µg in 400 µl), scales with complexity of panels and number of samples | **Automated IBEX** |
| **Software and system requirements for image processing**<br>• Imaris and Imaris File Converter (x64, version 9.5.0), variable depending on bulk pricing and bundling with other licenses, $5,000-10,000 per license<br>• Imaris Viewer $0<br>• Custom Imaris Extension $0<br>• System requirements for image processing, highly variable $4,000-30,000<br>    ○ MacBook Pro 2.3 GHz 8-Core i9 with 32 GB memory $4,000, scaled to multiple nodes on a high-performance cluster (>$5,000 per node) for batch processing<br>    ○ HP Z8 G4 Workstation or equivalent $16,000-21,000 | **Manual and Automated IBEX** |